\documentclass[11pt,a4paper]{article}
\usepackage{jheppub}

\usepackage{array}
\hypersetup{colorlinks=true,linkcolor=blue,citecolor=magenta,filecolor=magenta,
  urlcolor=cyan}

\usepackage{multirow}
\usepackage{float}
\usepackage{slashed}
\usepackage{tikz-feynman}
\usepackage{subfig}

\definecolor{darkmagenta}{rgb}{0.55, 0.0, 0.55}
\definecolor{blue}{rgb}{0.0, 0.5, 0.69}

\begin{document}

\title{Boosted decision trees in the era of new physics: a smuon analysis case study}

\author[a]{Alan S. Cornell}
\emailAdd{acornell@uj.ac.za}

\author[b]{\!\!, Wesley Doorsamy}
\emailAdd{wdoorsamy@uj.ac.za}

\author[c]{\!\!, Benjamin Fuks}
\emailAdd{fuks@lpthe.jussieu.fr}

\author[a]{\!\!, Gerhard Harmsen}
\emailAdd{gerhard.harmsen5@gmail.com}

\author[a,d]{\!\!, \!and Lara~Mason}
\emailAdd{mason@ipnl.in2p3.fr}

\affiliation[a]{Department of Physics, University of Johannesburg, PO Box 524,
  Auckland Park 2006, South Africa}
\affiliation[b]{Institute for Intelligent Systems, University of Johannesburg, PO Box 524,
  Auckland Park 2006, South Africa}
\affiliation[c]{Laboratoire de Physique Th\'eorique et Hautes Energies (LPTHE),
  UMR 7589, Sorbonne Universit\'e et CNRS, 4 place Jussieu,
  75252 Paris Cedex 05, France}
\affiliation[d]{Universit\'e de Lyon, F-69622 Lyon, France: Universit\'e Lyon 1,
  Villeurbanne CNRS/IN2P3, UMR5822, Institut de Physique des 2 Infinis de Lyon}
  
\date{\today}

\abstract{Machine learning algorithms are growing increasingly popular in particle phy\-sics analyses, where they are used for their ability to solve difficult classification and regression problems. While the tools are very powerful, they may often be under- or mis-utilised. In the following, we investigate the use of gradient boosting techniques as applicable to a generic particle physics problem. We use as an example a Beyond the Standard Model smuon collider analysis which applies to both current and future hadron colliders, and we compare our results to a traditional cut-and-count approach. In particular, we interrogate the use of metrics in imbalanced datasets which are characteristic of high energy physics problems, offering an alternative to the widely used area under the curve (\textit{auc}) metric through a novel use of the \textit{F-score} metric. We present an in-depth comparison of feature selection and investigation using a principal component analysis, Shapley values, and feature permutation methods in a way which we hope will be widely applicable to future particle physics analyses. Moreover, we show that a machine learning model can extend the 95~\% confidence level exclusions obtained in a traditional cut-and-count analysis, while potentially bypassing the need for complicated feature selections. Finally, we discuss the possibility of constructing a general machine learning model which is applicable to probe a two-dimensional mass plane. }

\maketitle
\flushbottom


\section{Introduction}
A key objective among experimental high energy physics programs probing both the Standard Model (SM) and Beyond the Standard Model (BSM) theories is to fully exploit the discovery potential of the data produced at colliders. To that end, machine learning techniques have been in use in high energy physics for 30 years~\cite{DENBY1988429, PhysRevLett.65.1321, LONNBLAD1991675}, and are becoming ever more integrated in multiple aspects of the particle physics universe as they evolve to allow for greater insight into existing and future data. In particular, the use of multivariate techniques in the analysis of data from colliders is growing. Since pioneering works that illustrated the use of boosted decision trees (BDTs) for particle identification~\cite{Roe:2004na, Yang_2005}, multivariate machine learning techniques have proven themselves to be a viable alternative to traditional methods.  It has previously been shown that BDT methods offer windows to new physics in taking a more sophisticated approach than standard cut-and-count methods. Indeed, recent experimental analyses in Refs.~\cite{Aaltonen:2010jr,  Aubert:2006av, Abazov:2006gd, 2012142, CMS-PAS-HIG-11-024} (and others) have shown the possibility of improving the signal significance through the use of decision tree-based tools, relying on the algorithm to identify the optimal way to analyse the events and to separate the signal under consideration from the related background.\ 

While it is arguable that BDTs have achieved a level of popularity in particle physics, we have found few resources to serve as a guide detailing how one should employ such algorithms. Broad reviews on the topic of machine learning in particle physics have been written~\cite{Albertsson:2018maf, PRESSCUT-H-2018-405, Carleo:2019ptp, Bourilkov_2019, Schwartz:2021ftp, Guest_2018, Larkoski_2020}, which give an overview of the many applications of machine learning tools in the field. There also exist reviews on the applications of neural networks~\cite{Brehmer:2018eca, Brehmer:2018kdj, Hollingsworth:2020kjg, Nachman:2019dol, Strong:2020mge}. In the case of BDTs, reviews do exist and discuss their utility in particle physics in general terms~\cite{Coadou:2013lca}. Our goal is therefore to complement these existing studies by providing a guide on the usage of a BDT algorithm using a BSM physics analysis as an example. In the following, we describe the entire process, from simulations to the machine learning analysis, and compare to an existing work which employs an advanced cut-and-count analysis. As is demonstrated in this work, while cut-and-count methods may require new high-level variables to be constructed in order to extract well-hidden signals, BDTs offer equally good performance training on only basic variables. We also examine a range of mass benchmarks, and discuss the applicability of one trained model on another benchmark. This offers an analysis of the utility of BDT models across a variety of scenarios, allowing for inclusive scans across parameter spaces. \

A high energy physics analysis at colliders traditionally consists of comparing real data obtained from a collider, such as the Large Hadron Collider (LHC) situated at CERN, Geneva, with simulated signal and background processes, to gauge the understanding of the results. The goal is to separate signal from background. The datasets are often largely imbalanced, and rare signal processes may mimic the distributions of relevant backgrounds. Additionally regions of feature space, which is a collection of features which characterise the data, often exhibit overlap from both distributions. A foundational method of analysis, `cut-and-count', applies thresholds on discriminating variables (cuts) on both signal and background, and counts the number of signal $s$ and background $b$ events that pass the selection. Once the optimal cuts have been obtained, the significance of the signal over the background can be measured using $s$ and $b$, where there is some uncertainty, $\Delta b$, associated to the background which includes statistical and systematic uncertainties. However, cut-and-count analyses are not sensitive to the regions of significant overlap, where either large amounts of signal must be cut, or background kept and counted. While the number of signal events $s$ also has associated to it some uncertainty, this is generically included in a quoted error band and is thus not explicitly accounted for here.  \

Instead of relying on cut-and-count techniques, one may employ machine learning to classify signal and background events. A generic supervised machine learning algorithm takes as input the observables of a given dataset and performs a regression (where the model learns a continuous function) or classification task. Outputs of a classification task are restricted to a finite set of values or classes~\cite{Bourilkov:2019yoi}, where the model learns to index the objects. In particular, supervised machine learning algorithms are trained on labelled simulated events, where traditionally signal processes are labelled with $1$ and background with $0$. During training, the algorithm learns the features of the samples and produces a model. The true measure of the generalisability of a model is its application to unseen and unlabelled test samples, which may be simulated events or real data. Decision trees extend a cut-based analysis by accommodating events which may fail a particular selection criterion. Many or most events do not have all characteristics of signal or background, and so we may not wish to reject every event that just fails one criterion. Instead, a decision tree may not reject events when they fail one criterion, but will consider whether other criteria may help classify them better. In this approach, which differs from the strict `cut-and-count' of traditional analyses, we may see improvements in hard-to-reach signal parameter space. \

A generic machine learning algorithm is particularly useful in applications to large datasets, such as those produced at particle colliders, where it can be utilised to find new features in data. In particular, this work focuses on the use of machine learning tools, and in particular BDTs, to separate signal processes of interest from background processes, where the signal is almost always far more scarce. We employ here the {\sc XGBoost} package~\cite{Chen:2016:XST:2939672.2939785} in addressing a classification problem typical of a particle physics analysis. We also address the use of performance metrics in an imbalanced case such as this, and show that alternatives to the widely used \textit{auc} (area under the receiver operating characteristic curve) should be considered. Instead, we compare the performance of three metrics, including the \textit{ams} metric, which quantifies the significance of signal over background~\cite{Cowan:2010js}. We motivate the use of the \textit{F-score} metric in place of the \textit{auc}, which proves to be more stable. We motivate that the most popular metrics, such as the \textit{auc}, are not always the most advantageous choice.\

As an example case, we examine the production of a pair of right-handed smuons decaying to a muon pair which could be produced at the LHC in supersymmetric scenarios~\cite{Haber:1984rc, NILLES19841, Fuks:2019iaj}. This BSM process is used as a case study, where it has previously been examined using a traditional cut-and-count analysis which additionally employed an advanced dynamic jet veto technique~\cite{Pascoli:2018heg, Pascoli:2018rsg} to separate signal from background. In order to obtain their results, the authors constructed sophisticated new variables on which to perform selections to eliminate background. It is a goal of this work to demonstrate that with the use of machine learning tools one may avoid the construction of new features, as the standard features may provide equally good results. In comparing the results of the \textit{auc} and $F$-score metrics, we assess the improvements at each benchmark that the machine learning algorithms can make over the cut-and-count approach. Additionally, we motivate the usefulness of a machine learning model which can be broadly applied across the mass parameter space in a given BSM theory.

 In this manuscript we begin in section~\ref{sec:theory} with a discussion of a generic machine learning algorithm and key concepts in an analysis such as this one, before turning our attention to the case study. We outline the layout of a generic machine learning algorithm, and further describe how a BDT is optimised through the minimisation of a loss function. We discuss the application of key machine learning concepts to particle physics, and introduce the metrics to be employed in the training and testing of our algorithm, concluding with a brief discussion of event weights. Next, in section~\ref{sec:generation} we outline the physics case at hand, describing sample event generation and the tools employed. We make use of the {\sc XGBoost} machine learning toolkit, and give a detailed analysis of the features of the samples in section~\ref{sec:featselec}. In particular, we investigate feature importances and the use of various performance metrics for model evaluation in the training and testing of multivariate techniques, where these avenues play key roles in our elucidation of the `black box' and decision-making process. We analyse the use of the \textit{auc}~\cite{rocaucref} metric in combination with the \textit{ams}~\cite{Cowan:2010js} metric, and compare performance to the combination of the \textit{F-score}~\cite{fscoreoriginal, Zhang2009} metric with the \textit{ams} metric. We conclude the section with a discussion of the hyperparameter optimisation employed prior to training the model. Finally, section~\ref{sec:results} outlines the results of the training and testing stages, including a discussion of the utility of simple and derived features, and concludes with an examination of the generalisability of a model trained on one benchmark to other benchmarks. Our discussion is focused on {\sc XGBoost}, but the principles are widely applicable, as we intend this paper to be a recipe for employing a BDT algorithm in an analysis, approached strictly from a physics perspective.\
 
\section{Machine learning in physics} 
\label{sec:theory}
In the following, we describe the functioning of a machine learning algorithm from an optimisation perspective, and outline a number of key definitions. A central goal in this analysis is to compare the utility of three of metrics, which are used during the training and testing stages of the multivariate selection, and to outline the capabilities of BDT algorithms.\

A supervised learning problem assumes there exists a mapping $f$, which may be stochastic or deterministic, relating elements of dataset $\mathcal{D}$,
\begin{equation}
y = f(x), \quad \mathcal{D} = \{ x_i, y_i \},
\end{equation}
where $x_i$ are the input data, which have associated to them some features, and $y_i$ the final outcome to be predicted, or the target. The supervised learning goal for a classification problem is to approximate $f$ given $\mathcal{D}$ using the characteristics of the objects and the classification as the target, returning an approximation to $f$:
\begin{equation}
\hat{f} = \mathcal{A}(\mathcal{D})
\end{equation}
which may then be used to categorise or classify unseen and unlabelled inputs. The algorithm finds this mapping by solving an optimisation task, measuring the quality of the prediction for a single object $x_i$ with a loss function $L(y_i,\hat{f}(x_i))$ which should be minimised to find the best model according to some metric. The division of simulated data into training and testing sets allows for the assessment of the quality of the model, where the training set is used to fit and tune the model, and the test set to measure the performance of the model. Finally, a validation subset is typically used to measure the accuracy of the model during its tuning.\

Machine learning algorithms are subject to underfitting and overfitting, which should be carefully controlled. The first occurs in a model which is unable to fit even the training set due to low complexity, and the latter when the model is too complex, learning the intricacies of the training set too well such that the algorithm is unable to generalise. If a large difference between the quality metric on the training and test samples emerges, the model is likely overfitting. The behaviour and complexity of a model is controlled by the bias-variance trade-off, where underfitting and overfitting are finely balanced. A well-tuned model would be sophisticated enough to interpret and express underlying patterns in data, but not so complex that it reproduces spurious structures.

\subsection{Boosted decision trees}
BDT multivariate techniques are a family of supervised machine learning tools whose robust classification power~\cite{Roe:2004na} makes them a popular choice in high energy physics applications. While state of the art deep learning with neural networks perform well in particle physics scenarios, they often require large amounts of data for training and are CPU-intensive~\cite{lalchand2020extracting}. In contrast, decision trees are robust machine learning algorithms, favoured by many due to their ability to deal well with missing values, imbalanced datasets, and redundant attributes~\cite{4e602701df284c70b06d479d4100c1d7}. These characteristics, while signal-dependent, often appear in a generic dataset produced by the particle collider such as the LHC. For instance, features such as the angular separation $\Delta R\left(\ell_1,\ell_2\right)$ between two leptons $\ell_1$ and $\ell_2$ would be missing in an event where only one good quality lepton was detected. The ability to deal with these characteristics makes decision tree-based algorithms a worthy candidate to tackle these problems.\

Simple decision tree-based algorithms featuring single trees have been in use for many years since first appearing in the 1980s~\cite{cart84} but are known to be unstable alone, and so are referred to as \textit{weak learners} because they do not perform well as standalone classifiers. `Boosting' improves on single decision trees by combining numerous weak learners, resulting in a powerful multivariate algorithm. Boosted algorithms are built on the idea that while it is hard to make a good learner, it is easy to make many weak learners, which are built successively to improve on the failures of the predecessors. Rather than relying on a single tree, boosting algorithms iteratively construct new base learners (trees) maximally inversely correlated with the gradient of the loss function of the entire ensemble of learners~\cite{natekingb}, and the discriminating model utilises the weighted average score from many decision trees. At each step, the events which are misclassified (signal as background, or \textit{vice versa}), are given a larger weight (not to be confused with the weights present in high energy physics simulations), or \textit{boosted}, and the a new tree is built on the new weights. As such, learners which are added to the model later are better at classifying samples which were previously misclassified. In this way, the power of boosting lies not with the strength of the final tree (which would be expected to have very undesirable performance on the full dataset) but rather on the ensemble of all the week learners at once. In this work, we will consider binary trees, which output only two possible classes: $1$ or $0$.  \

A boosting algorithm exploits a large number of weak learners $h_k$ where each improves the mistakes of the previous learner, creating a general meta-algorithm which can create a strong hypothesis from sum of weak learners with associated weights $\alpha_k$~\cite{Coadou:2013lca}
\begin{equation}
F(x) = \sum_{k=1}^{N_{tree}} \alpha_k h_k(x),
\label{eq:summedclass}
\end{equation}
where $N_{tree}$ denotes the number of decision trees built as classifiers, and $x$ is the feature space of the event.
In this way, the resulting model is a weighted sum of the weak learners' predictions. The weak learner decision trees are a collection of nodes; the internal nodes (containing predicates) and the leaves. The predicates, or splitting rules, check the value of a given variable to be used for classification and compare it to some threshold. Simply put, the nodes take an input $x$ and return $0$ or $1$ if the predicate is respectively true or false in comparison to a threshold $t$ (for example, $f_1^x < t$, where $f_1^x$ is a feature associated to input $x$). At the end of the chains of nodes are leaves, which allow the model to make predictions; an object moves down the chain, being assigned a direction at each node, and eventually landing on a leaf which classifies it. The splitting at each node should maximise the purity of the split or the separability of signal from background.\

A general boosting process with a generic loss function ${L}(y_i,\hat{f}(x_i))$ proceeds by iteratively finding the next learner $h_i(x)$ to be added to the ensemble of decision trees for a dataset of size $N$. This is done through minimisation of the loss function
\begin{equation}
\sum_{i=1}^{N}{L}_n\left(y_i,~F_{n-1}(x_i) + \alpha_n h_n(x_i)\right) \rightarrow \text{min}_{\alpha,h},
\label{eq:LF}
\end{equation}
by finding appropriate values for the weight $\alpha_n$ associated to the new learner $h_n(x_i)$, where $F_{N-1}(x)$ is the function built up to that point. \

Gradient boosting machine learning algorithms utilise gradient descent~\cite{Friedman00greedyfunction} as a loss function, adding a learner on each iteration, and is a general approach to boosting, adding weak learners that approximate the gradient of the loss function. In order to utilise gradient descent, the gradient of the loss function is computed with respect to its predictions, using the residuals of the model, $y_i - F(x_i)$, which are the elements of training data where the model had the largest errors. The role of the next learner added is to compensate for the shortcomings of the existing model. In essence, the loss function to be minimised is simply a function of two variables: the true labels and the predicted labels. \

Gradient boosting models achieve state of the art performance, generally outperforming linear machine learning models~\cite{Friedman00greedyfunction}. The {\sc XGBoost}~\cite{Chen:2016:XST:2939672.2939785} machine learning algorithm for supervised learning is an extension of generic gradient boosting methods, employing the `extreme' gradient boosting method, designed for scalability and performance, and is our choice for this study. The aim is to estimate the gradient direction of the differentiable convex loss function~\cite{Chen:2016:XST:2939672.2939785} using second-order derivatives, allowing for the extraction of more information to fit the learners. In particular, along with employing gradient boosting with second-order optimisation, {\sc XGBoost} utilises penalised loss and a particular choice of impurity criterion which further differentiates it from gradient boosting algorithms, which utilise first-order derivatives. We refer the interested reader seeking further details on boosting and gradient descent methods to Ref.~\cite{schapireboosting}.

\subsection{Metrics and class imbalance}

A characteristic of high energy physics datasets, including real data from colliders and simulated events, is a large class imbalance. Typical datasets feature many more background events than signal events, where the background events may come from a number of processes which mimic the signal of interest. The signal can have many incarnations according to the model parameters, but which finally results in the same `signature' which is characteristic of the process of interest. In order to measure the success of a given machine learning model, a metric should be introduced which quantifies the model's ability to correctly classify the signal and background events. The measure of classification performance is needed both for assessing the final result and for tuning the hyperparameters during training. A simple metric to consider, and one that is widely used, is the accuracy metric, which returns the percentage of correctly identified events. However, class imbalance can exert a major impact on the accuracy metric, modifying its meaning and reliability. In particular, excellent accuracy may be obtained by classifiers simply guessing the majority (background) class~\cite{Chawla2005DataMF}. The simplicity of the accuracy metric leads to a bias towards the majority class, leading to the metric being less discriminatory and less informative~\cite{Hossin2015}. The imbalanced dataset problem has been studied in the machine learning community, with many looking to improve on the performance of the accuracy metric~\cite{Chawla2005DataMF, Hossin2015, saito2015, DBLP:journals/corr/abs-1106-4557, LUQUE2019216}.\

A popular choice of metric amongst the high energy physics community is the \textit{area under the receiver operating characteristics curve (auc)}, which reflects the overall ranking performance of a classifier~\cite{Hossin2015}.  This metric considers the true and false positive identification rates. This metric is a standard metric for use in machine learning for particle physics, and has shown reasonable performance for imbalanced datasets~\cite{10.1007/978-3-642-04962-0_53}.  As an alternative metric, the \textit{F-score}~\cite{DBLP:journals/corr/LewisG94} may be considered to evaluate the classification power of a model. The \textit{F-score} metric is a measure of accuracy, but has been reported to perform better as a discriminator than the accuracy metric, which simply measures the number of predictions the model made correctly~\cite{Hossin2015}. It is a popular metric for imbalanced datasets~\cite{4630300621} but is not usually considered in high energy physics analyses, and may improve on the \textit{auc} performance~\cite{10.1145/1143844.1143874}. Finally, the \textit{ams} metric is equivalent to the significance used in traditional cut-and-count analyses, measuring the discovery potential of an analysis. It is calculated using a formula for the discovery significance for Poisson data. In the machine learning context, and particularly with very imbalanced datasets, it is too volatile to be used in training or hyperparameter selection, but may be used to evaluate the model at the final stage given the number of true background and signal events selected as signal.\

The difficulties introduced by large class imbalances may be handled by a careful choice of metric, or by altering the size of the training set by rescaling of the weights (discussed in section~\ref{sec:weights}) to balance the classes. However, testing must take place on the true size of classes to represent real world performance, and test results often suffer from the sheer difference in numbers between signal and background. The advantage of rescaling weights in training is metric-dependent, but the \textit{auc} metric performs well on balanced training classes~\cite{DBLP:journals/corr/abs-1106-4557}. The accuracy is not improved through such a method. Additional methods of sampling will not be discussed here.\

In the following, we further outline the characteristics of each metric (the popular \textit{auc}, the alternative \textit{F-score}, and the \textit{ams}) in preparation for their comparison in later sections. In all cases, metrics used on imbalanced datasets should be implemented carefully, and it is often prudent to compare metric performance to identify falsely inflated (or deflated) results. 

\subsubsection{The \textit{auc} metric}
The \textit{auc} is controlled by the true and positive rates, and varies between results of $1$ for a model which perfectly separates signal from background and $0.5$ for no improvement over random guessing. The better the true positive rate the higher the \textit{auc} score. During a classification, the correct and incorrect predictions by a model are stored and usually displayed in a confusion matrix, as displayed in table~\ref{tab:cm}, and on which the \textit{auc} and $F-1$ score metrics are based. Predictions made by the model are grouped into `true positive (TP)' and `true negative (TN)', for correctly classified signal events and background events respectively, or `false negative (FN)' and `false positive (FP)', for incorrectly classified signal events and background events. 

\begin{table}
\centering
 \renewcommand\arraystretch{1.35}
 \setlength\tabcolsep{6pt}
\begin{tabular}{cc|cc|c}
& & \multicolumn{2}{c}{\textit{Predicted classification}} & \\
     &    & Negative & Positive & \\
 \hline        
\multirow{2}{*}{\textit{True classification}} &  Negative & TN & FP & CN \\
 & Positive & FN & TP & CP \\
 \hline
 &          & RN & RP & N \\
\end{tabular}
\caption{A typical confusion matrix, where TN/FN indicates true negative/false negative, and TP/FP indicates true positive/false positive. CP/CN is the number of events in the positive/negative class, and RP/RN is the number of predicted positive/predicted negative events.}
\label{tab:cm}
\end{table}

Plotting the TP rate (TPR) against the FN rate (FNR) for a number of candidate threshold values to find the receiver operating curve (\textit{roc}) curve, the \textit{auc} is the area under this curve and is a measure of separability for our predictions, where
\begin{equation}
\begin{split}
    & \text{TPR} \equiv \frac{\text{TP}}{\text{TP} + \text{FN}}\\
    & \text{FPR} \equiv \frac{\text{FP}}{\text{TN}+ \text{FP}}.
    \end{split}
\end{equation}

\subsubsection{The \textit{F-score} metric}
The \textit{F-score} metric is an alternative to the \textit{auc}, and is also defined by values from the confusion matrix. The metric is a weighted mean of the precision and recall, where the precision is the ratio or true positives (events correctly identified as signal) to all positive results (both correctly and incorrectly identified), and the recall is the ratio of true positives to the total number of true signal events, such that 
\begin{equation}
\begin{split}
    \text{precision} & = \frac{\text{TP}}{\text{TP} + \text{FP}},\\
  \text{recall} & = \frac{\text{TP}}{\text{TP} + \text{FN}}.
  \end{split}
\end{equation}
The recall is defined identically to the TPR of the previous subsection. The \textit{F-score} metric is defined in a general manner according to a positive real parameter $\beta$, which defines the ratio of importance between the precision and recall,
\begin{equation}
    F_{\beta} = \left(1 + \beta^{2} \right) \frac{\text{precision}  \times \text{recall} }{\left(\beta^{2} \times \text{precision} \right) + \text{recall}}.
\end{equation}
The \textit{F-score} metric holds the particular advantage that it may be tuned, via the $\beta$ parameter, to tighten or loosen selection criteria. In particular, for $\beta = 1$, we define the \textit{F-1 score metric}, where precision and recall are equally important. One may also consider an unweighted \textit{F-score} by modifying the value of $\beta$ to emphasise either the precision or recall, where higher precision imposes stricter selections for a positive result, and higher recall allows more events to be identified as positive. While the former will lead to a very pure sample of positively identified events, the latter will reduce the number of false negatives, which may be crucial in a highly imbalanced particle physics dataset with very few true signal events.\

A large area under the precision-recall curve (the plot of the precision versus the recall for varying probability thresholds) indicates a high recall and high precision rate, or low false-negative and false-positive rates. Since precision is directly influenced by the class imbalance, precision-recall curves may be better indicators of the performance of the algorithm in cases often seen in high energy physics.\

\subsubsection{The \textit{ams} score}

In a particle physics analysis, the quality of the identification of an event as a signal or background event is measured by the significance which may be denoted in a number of ways, such as $s/\sqrt{s+b}$ or $s/\sqrt{b}$~\cite{Read:2002hq}. As previously, $s$ is the number of true signal events and $b$ the number of background events which pass the signal selection criteria. In a standard counting experiment of real data in an area where we are looking to disprove the background-only hypothesis, one finds a Poisson distribution with $n$ events, with a mean of $s+b$. The discovery significance quantifies the significance with which the background-only hypothesis is rejected. In planning a statistical analysis in a particle physics setting, the aim is to maximise the discovery significance for a given signal on the assumption that the signal is present~\cite{pmlr-v42-cowa14}. In using {\sc XGBoost}, we utilise the approximate median discovery significance
(\textit{ams}) in estimating the sensitivity of the analysis to our signal. It is
defined by~\cite{Cowan:2010js}
\begin{equation}
  \textit{ams} = \sqrt{2\left( \left( s + b \right) \ln \left( 1 + \frac{s}{b}\right)-s \right)},
  \label{eq:ams}
\end{equation}
and emerges as the expectation of the significance for a counting experiment with $n$ counts following a Poisson distribution featuring $n$ events. In particular, this relates to a counting experiment with a known value for the background $b$. In a general particle physics analysis of this nature, the background is generally associated to an uncertainty $\Delta b$, and the Asimov significance is defined by
\begin{equation}
  \sqrt{2\left( \left( s + b \right) \ln \left( \frac{(s+b)(b+\Delta b^2)}{b^2 + (s+b)\Delta b^2} \right) - \frac{b^2}{\Delta b^2}\ln \left(1+\frac{\Delta b^2 s}{b(b+\Delta b^2)} \right)\right) }.
  \label{eq:ams2}
\end{equation}
 In a particle physics context, a significance of about $2\sigma$ is considered acceptable for exclusion, $3\sigma$ for hints of a discovery and $5\sigma$ for a full discovery. The \textit{ams} represents the significance, and therefore discovery potential, of a given analysis. It has been argued that the instability of the \textit{ams} makes it an unsuitable choice as the optimisation metric for reweighted events during training, although it has been employed as a direct optimisation metric with some success~\cite{pmlr-v42-cowa14}. As an alternative, a model may be trained using a metric such as \textit{auc}, after which the \textit{ams} on the trained model can be evaluated.\

\subsection{Weights}
\label{sec:weights}
In a high energy physics dataset, simulated events almost always have associated to them a weight, which indicates that a given event in the sample usually corresponds to more events in reality, although in some weighting cases it corresponds to fewer. Weights may in principle be arbitrarily small, leading to large demands on CPU. That is, the sum of weights of a dataset will generally sum to the cross section of the process at hand. For use in our case, where the weights are utilised by the \textit{ams} metric, we must have that the sum of weights corresponds to the number of expected events. As such, the formula 
\begin{equation}
N_{\rm events} = \mathcal{L} \times \sigma
\label{eq:nevents}
\end{equation}
should be used as appropriate. Here $\mathcal{L}$ is the integrated luminosity, which measures the rate of collisions produced in a detector. Given this sum of weights, the purity of the dataset can be defined as
\begin{equation}
p = \frac{s}{s+(b+\Delta b)} = \frac{\sum_i w_i^s}{\sum_i w_i^s + \sum_j (w_j^b+ \Delta w_j^b) },
\end{equation}
for $i$ signal events with weights $w_i$,  and $j$ background events with weights $w_j$, where the background events are associated to an uncertainty $\Delta b$ due to the shape of their distribution. \

The $ams$ metric is normalised by the sum of weights of the data on which it is evaluated, where this sum should be the number of expected events, as in eq.~(\ref{eq:nevents}). This implies that weights must be handled carefully. If weights are arbitrarily rescaled, the $ams$ value computed on them will be unrealistic. For any training, testing, and validation subset, the sum of the weights must be the same in order to obtain meaningful $ams$ values. When analysing $x\%$ of the dataset, weights should be rescaled by $1/x\%$. Failure to do this may lead to lower values of the $ams$ score.\

It should be clear that these weights are associated to the simulation of physics processes, and should be handled with care. They are not related in any way to the weights $\alpha_k$ with which weak learners $T_k$ are added to make a classifier as in eq.~(\ref{eq:summedclass}).

\begin{table}
 \renewcommand\arraystretch{1.35}
 \setlength\tabcolsep{6pt}
\centering
\begin{tabular}{c|c|c}
Quantity & Physics meaning & Machine learning application \\
\hline
\hline
$y = f(x)$ & Signal or background & Label $(0,1)$ for $(b,s)$.\\
\hline
\multirow{2}{*}{$\mathcal{D}(x_i, y_i)$} & The set of simulated events & \multirow{2}{*}{The training dataset}\\
 & $x_i$ including the label $y_i$ & \\
\hline
\multirow{3}{*}{${ L}(y_i,\hat{f}(x_i))$} & Discrepancy between   & Loss function to be iteratively  \\
& predicted labels  & minimised: penalise physically \\
& and true labels & inconsistent outcomes  \\
\hline
\hline
TPR & Correctly classified  & \multirow{2}{*}{Signal efficiency} \\
             &   signal events \\
\hline
FPR & Incorrectly classified  & \multirow{2}{*}{Background efficiency} \\
& background events & \\
\hline 
 \multirow{4}{*}{$auc$} &  Metric which is   & Probability for a classifier to rank a  \\
 &maximised with ability    & randomly chosen signal event  \\
 &to classify signal and & higher than a randomly chosen \\
 &  background & background event \\
\hline
\multirow{3}{*}{$ams$} & \multirow{3}{*}{Discovery potential of signal } & Expectation value of  \\
& & significance $Z$ assuming  \\
& & a Poisson distribution of events \\
\hline
Node & Assessment of event  & \multirow{2}{*}{Predicate comparing to threshold}\\
& given a cut on a variable & \\
\hline
Leaf & $1$ or $0$ classification & Termination of the analysis of a single event\\
\end{tabular}
\caption{A dictionary of the terms and symbols appearing in this paper, including their context in both physics and machine-learning studies.}
\label{tab:concepts}
\end{table}

In table~\ref{tab:concepts}, we summarise this section with a list of quantities and jargon which appear often in such a scenario, indicating both the physics implication and machine learning application for each term. 

 \section{An illustrative example}
\label{sec:generation}
In the following, we examine the application of the {\sc XGBoost} BDT machine learning algorithm to a BSM example case featuring smuons and neutralinos originating from weak-scale supersymmetry~\cite{Haber:1984rc, NILLES19841, Fuks:2019iaj}. Such models are popular for their ability to address the hierarchy problem of the SM, dark matter, and gauge unification, amongst other reasons. Sleptons are not yet excluded at the LHC above several hundred GeV for neutralinos masses around $100~{\rm GeV} - 300~{\rm GeV}$~\cite{ATLAS:2019lff, CMS:2020bfa}. We focus here on the simplified case where coloured particles are decoupled. We begin by describing the simulation of background and signal data, and then outline methods for feature investigation and selection. Following this, we proceed with a hyperparameter optimisation strategy. These feature and hyperparameter investigations are all completed before the {\sc XGBoost} toolkit is trained, but form a vital part of tuning and preparing the machine learning model. 

In the following, the simulated samples were divided into a 30\% test set and a 70\%  training set. The test set was witheld during both the hyperparameter selections and model training process, and only employed after all training and tuning had been completed, ensuring that the performance on the testing set remained unbiased. The test sample is then used to assess the performance of the  {\sc XGBoost}  on unseen samples in section~\ref{sec:results}. For this classification task we used the binary-logistic loss function to train the {\sc XGBoost} model. This loss function is given as \cite{NielsenNN2015} 
\begin{equation}
    L(y_{i},\hat{f}(x_{i})) = -\frac{1}{N}\sum_{i=1}^{N}y_{i}\text{log}(\hat{f}(x_{i})) + (1-y_{i})\text{log}(1-\hat{f}(x_{i})),
\end{equation}
where N is the number of samples, $y_{i}$ is the label of the $i$th data point and $\hat{f}(x_{i})$ is the probability of $x_i$ being either a 0 or a 1. 

\subsection{The physics case}

\begin{figure}
\centering
\includegraphics[width=0.4\textwidth]{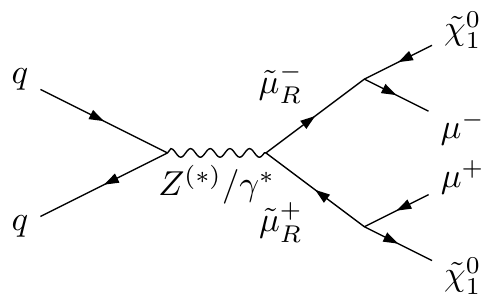}
\caption{Representative Feynman diagram for the process considered, featuring the production of a muon pair and a neutralino pair in a proton-proton collision.}
\label{fig:feyn}
\end{figure}

We begin by outlining the signal and background processes relevant to this study. The partonic process of interest is the Drell-Yan production of a smuon pair, decaying to neutralinos and a pair of opposite sign muons in the context of the Minimal Supersymmetric Standard Model,
\begin{equation}
pp\rightarrow \gamma^*/Z^{(*)} \rightarrow \tilde{\mu}_R^+ ~\tilde{\mu}_R^- \rightarrow \mu^+\mu^-\tilde{\chi}^0_1 \tilde{\chi}^0_1.
\end{equation}
The process is illustrated in the Feynman diagram presented in Figure~\ref{fig:feyn}. We consider a simplified model in which all other superpartners are decoupled. The Lagrangian of interest is
\begin{equation}
\begin{split}
\mathcal{L} = & \big[ D_\mu\tilde{\mu}_R^\dagger\big]\big[ D^\mu\tilde{\mu}_R\big] + \frac{i}{2}\overline{\tilde{\chi}^0_1}\slashed{\partial}\tilde{\chi}^0_1 - m^2_{\tilde{\mu}_R}\tilde{\mu}_R^\dagger \tilde{\mu}_R - \frac{1}{2}m_{\tilde{\chi}^0_1}\overline{\tilde{\chi}^0_1}\tilde{\chi}^0_1  \\
&   -\frac{\sqrt{2}e}{c_W}\left[ \left( \overline{\tilde{\chi}_1} P_R\mu \right) \tilde{\mu}_R^\dagger + \text{h.c.}\right],
\label{eq:lag}
\end{split}
\end{equation}
where $c_W$ is the cosine and sine of the electroweak mixing angle, $e$ is the electromagnetic coupling constant, and $P_R$ is the right-handed chirality projection operator. The neutralino $\tilde{\chi}_1$ of mass $m_{\tilde{\chi}_1}$ is taken as bino-like. The masses of the smuons $(m_{\tilde{\mu}})$ and neutralinos $(m_{\tilde{\chi}})$ are left as free parameters, with $m_{\tilde{\chi}} \in [55,500]~{\rm GeV}$ and $m_{\tilde{\mu}} \in [200,510]~{\rm GeV}$. The lower bound on the smuon mass reflects the range of sensitivity for the LHC, guaranteeing that final state objects possess enough energy to be of good quality in the detector, while the upper bound restricts events to those possessing large enough cross sections for reasonable event generation~\cite{Fuks:2013lya, Fuks:2013vua, Bozzi:2007qr}. Within these ranges, several benchmarks of interest have been identified. The benchmarks cover a number of regimes including \textit{compressed} scenarios, where the masses of the BSM particles are very close together, \textit{separated} scenarios, where the masses differ greatly, and \textit{neutral} scenarios, which lie in between the two. We have therefore grouped the benchmark points according to their mass ratios, $r = m_{\tilde{\mu}}/m_{\tilde{\chi}}$. We classify points with $r< 1.5$ as \textit{compressed}, points with $r\in [1.5,3]$ as \textit{neutral} and points with $r>3$ as \textit{separated}. We follow this approach as it is expected that the kinematics in these three scenarios will differ significantly, and it may be necessary to treat them separately. As this is a BSM physics scenario and the masses are not well constrained, we would like to investigate whether we are able to group the mass regimes and minimise the number of machine learning models to be trained. That is, if we train such a model on one benchmark in a given mass regime, we investigate whether the model can be applied to other benchmarks in the same regime with equally good results, or indeed whether it can be applied to other mass regimes.

\subsection{Event generation and tools}
Training and testing of the machine learning toolkit will be done on simulated samples, including both background and signal processes. The {\sc MadGraph5\_aMC@NLO (MG5\_aMC)}
framework~\cite{madgraph} has been used for the simulation at colliders for both the signal and background processes, where the decays of the smuons are handled by {\sc MadSpin}~\cite{Artoisenet:2012st} and {\sc MadWidth}~\cite{Alwall:2014bza}. The UFO model files~\cite{Degrande:2011ua} for use with {\sc MG5\_aMC} were created using the {\sc FeynRules}~\cite{Alloul:2013bka,Christensen:2009jx} package and are publicly available~\cite{Duhr:2011se}. In each case, the leading order set of NNPDF 2.3 parton densities
\texttt{NNPDF23\_lo\_as\_0130\_qed}~\cite{Ball:2013hta} is convoluted with the partonic cross section. In the case of the signal simulations using {\sc MG5\_aMC}, additional parton-level jets are allowed at the matrix element level, and an MLM merging procedure is employed to avoid double counting after matching with parton showering~\cite{Mangano:2006rw, Alwall:2008qv}. Relevant parameters for the jet merging procedure are included in table~\ref{tab:xqcut}.\

The simulation of collision events and reconstruction has been done in conjunction with {\sc Pythia}~8~\cite{Sjostrand:2007gs} to describe parton
showering and hadronisation, and with the {\sc Delphes}~3~\cite{deFavereau:2013fsa} software package that includes the anti-$k_T$ algorithm~\cite{Cacciari:2008gp} as implemented in {\sc FastJet}~3~\cite{Cacciari:2011ma} for event reconstruction. The default {\sc Delphes} card for the CMS detector was utilised, which is included in the standard  {\sc MG5\_aMC} installation, and proton collisions simulated at a centre of mass energy of $\sqrt{s} = 13\ {\rm TeV}$.\

A number of background processes are relevant to the considered final state, where SM processes featuring a final state pair of opposite sign leptons and neutrinos constitute the dominant background. Of particular relevance are the $t\bar{t}$ processes where both top quarks decay leptonically,
\begin{equation}
p ~p \rightarrow t~\bar{t} + X\rightarrow \ell^+ \ell^- b ~\bar{b} ~\nu_\ell ~ \bar{\nu}_{\ell} + X,
\end{equation}
and $W^+W^-$ events 
\begin{equation}
 p ~p  \rightarrow W^+ W^- + X \rightarrow \ell^+ \ell^- \bar{\nu}_{\ell} ~ \nu_\ell + X,
\end{equation}
where $X$ denotes QCD and QED radiation. Simple generator-level preselections are applied during simulation. Jets are required to have a transverse momentum greater than $20~{\rm GeV}$ ($p_T(j) > 20~{\rm GeV}$), and for leptons we require $p_T(\ell) > 10~{\rm GeV}$. Additionally, we require $|\eta| < 2.5$ for jets and leptons, and the minimum angular separation $\Delta R$ between jets and leptons is required to be $0.4$. The jet merging parameters for the signal events are presented in table~\ref{tab:xqcut}, and are constant across all signal samples.

\begin{table}
\centering
 \renewcommand\arraystretch{1.55}
 \setlength\tabcolsep{8pt}
\begin{tabular}{c||c|c|c|c}
Sample & $\ell\ell\nu\nu$ & $W^+W^-$ & $t\bar{t}$ & Signal\\
\hline
Q$^{\rm match}$ & 37.5~${\rm GeV}$ & 37.5~${\rm GeV}$ & 45~${\rm GeV}$ & 82.5~${\rm GeV}$ \\
\end{tabular}
\caption{Parameters for the multi-parton matrix element merging procedure, including the scale for background samples and for signal samples.}
\label{tab:xqcut}
\end{table}

At reconstruction level, further preselections are imposed. Each jet within an event, be it signal or background, is required to have $p_T > 25\ {\rm GeV}$, and each lepton $p_T > 10\ {\rm GeV}$. Again, the pseudorapidities of all accepted objects should satisfy $|\eta| < 2.5$. Further cuts are made for object overlap removal; for $\Delta R(j,\ell) < 0.1$, the jet in question is removed. The angular separation between leptons and the remaining jets is again assessed, and for $\Delta R(j,\ell) < 0.4$ the lepton is removed.  These selections are characteristic of ATLAS LHC analyses such as those detailed in Ref.~\cite{ATLAS:2019gdh}.\

We then place several additional baseline preselections on our events, following the strategy of Ref.~\cite{Fuks:2019iaj}. In particular, we require that all events possess one pair of opposite sign muons. Additionally, low mass hadronic resonances are rejected through requiring $m_{\mu\mu} > 20~{\rm GeV}$, where $m_{\mu\mu}$ denotes the invariant mass of the muon pair. Finally, the most significant removal of the background is achieved through a `stransverse' mass cut $M_{T2} > 90~{\rm GeV}$~\cite{Lester:1999tx, Cheng:2008hk}.

A goal of this work is to compare to the cut-and-count analysis done in Ref.~\cite{Fuks:2019iaj} and investigate any differences in feature handling that may appear in a BDT analysis of the same process. In the analysis to be compared to, the authors constructed sophisticated ratios, presented in eq.~\eqref{eq:ratios}, as innovative selection procedures to remove background samples while preserving signal events. These ratios are utilised as dynamic jet vetoes which  define and compare the leptonic activity and hadronic activity of the event~\cite{Pascoli:2018rsg, Pascoli:2018heg}. We aimed to test whether a BDT algorithm could compete with a cut-and-count analysis by extending the 95~\% confidence limit previously obtained, and in particular, whether the sophisticated ratios would again prove to be the most valuable. A generic BSM model features a wide range of possible masses for one or more new states, as is the case here for the smuon and neutralino. It was of interest to determine whether a machine learning model, trained on a single benchmark, could perform well on other mass benchmarks, as this offers ease in scanning across the parameter space.\

We have included in our feature space the six dynamic jet veto ratios which were identified as being important in the selection of signal processes, in order to assess their utility to a multivariate algorithm. The ratios feature a combination of transverse momenta and hadronic and leptonic summed quantities,
\begin{equation}
\begin{split}
& {\rm (a)}~p_T^{l_1}/p_T^{j_1}, \qquad {\rm (b)}~p_T^{l_1}/H_T, \qquad {\rm (c)}~S_T/p_T^{j_1}\\
& {\rm (d)}~S_T/H_T, \qquad {\rm (e)}~p_T^{l_2}/p_T^{j_1}, \qquad {\rm (f)}~p_T^{l_2}/H_T.
\end{split}
\label{eq:ratios}
\end{equation}
Here, the inclusive scalar sum of the $p_T$ of all the hadron clusters in an event is defined as
\begin{equation}
H_T \equiv \sum_{k\in \{ {\rm clusters} \} }{p}_T \left( {\rm cluster}_k \right)
\end{equation}
for jets with pseudorapidity $|\eta |\leq 4.5$, and the exclusive scalar sum of the $p_T$ of the two leptons with the highest $p_T$ is given as 
\begin{equation}
S_T \equiv \sum_{k=1}^2 p_T\left({\ell_k}\right).
\end{equation}
These features were identified as being necessary to achieve good performance in the cut-and-count analysis. In the following, we will consider training models both with and without these ratios. While in any machine learning analysis a good understanding of the data and the variables in the sample are crucial, to ensure sensible handling of the training and testing, we would like to investigate whether a BDT approach can avoid the need for feature engineering to the same extent. 

\subsection{Feature analysis and selection}
\label{sec:featselec}
In a generic model building scenario employing machine learning techniques in a high energy physics context, it is useful to perform exploratory data analysis by investigating the features which will be used by the machine learning model. By verifying how well the features describe the target and checking correlations between features, we are able to better understand the underlying physics as well as the plausibility of the learning structure, and avoid a `black-box', where through investigating which features are most important to the machine learning algorithm we can assess the advantage of the decision tree-based approach over cut-and-count methods. In the case that the algorithm does not perform well on primary features, more advanced feature construction may be performed. With the goal of understanding the feature space in mind, a two-step feature investigation was performed; an investigation into feature dimensionality and relevance was performed using a principal component analysis (PCA)~\cite{Jolliffe2011}, followed by feature selection utilising Shapley values~\cite{shapley} and feature permutation importance investigation. The PCA is an unsupervised technique which is widely used for dimensionality reduction, but in this case is utilised to investigate the variance of the feature space. The {\sc XGBoost} package features a supervised machine learning technique, but the use of the unsupervised PCA (where signal and background labels are not considered) allows for general consideration of the behaviour of the features. In contrast, feature permutation and Shapley investigations allow for the selection of the features which contribute most heavily to the supervised training. The first estimates the predictive value of a feature by measuring decrease in total score of the model when the feature is removed. Conversely, estimation of feature importance using Shapley values calculate the marginal contribution of the feature value across all the possible combinations of features. It has recently been demonstrated that Shapley values associated with BDTs have the potential to unravel the complicated $b\bar b h$ signal from the background~\cite{Grojean:2020ech}.\ 

We aim not only to determine how many features should be included in the machine learning model but also what the relative importance of the features are, and to better understand the structure of the model in feature space. The full list of features is given in table~\ref{tab:features}, where they are divided into primary and derived features. Also included in the table are a number of additional features employed in the analysis, such as the sum of the transverse momentum of all object candidates, $\sum (p_T)$, and the total missing energy $\slashed{E}_T$. It should be noted that decision trees do not require the same level of feature selection as other machine learning methods, as they are not significantly affected by a large number of variables, nor are they vulnerable to duplicate variables. They are able to handle discrete and continuous variables simultaneously, and are insensitive to the order in which the variables are introduced~\cite{Coadou:2013lca}. Instead, the dominant gain in feature selection arises in CPU improvements. It is nonetheless crucial to understand well the features included in the dataset to avoid a black-box implementation.\

\begin{table}
 \renewcommand\arraystretch{1.5}
 \setlength\tabcolsep{8pt}
\centering
\begin{tabular}{c||c}
 \multicolumn{2}{c}{Primary} \\
 \hline
Variable & Grouping \\
\hline
\hline
$n_{\ell}$  & \multirow{2}{*}{Final state: counting}\\
$n_{j}$ & \\
\cline{0-1}
$p_T^{j_1}$ & \multirow{2}{*}{Jet: momentum}  \\
$p_T^{j_2}$ &   \\
\cline{0-1}
$\eta^{j_1}$ &  \multirow{2}{*}{Jet: angular} \\
$\eta^{j_2}$ &    \\
$\phi^{j_1}$ &   \\
$\phi^{j_2}$  &  \\
\cline{0-1}
$p_T^{\ell_1}$  & \multirow{2}{*}{Lepton: momentum} \\
$p_T^{\ell_2}$  &   \\
\cline{0-1}
$\eta^{\ell_1}$  & \multirow{4}{*}{Lepton: angular} \\
$\eta^{\ell_2}$  &  \\
$\phi^{\ell_1}$ &   \\
$\phi^{\ell_2}$  &  \\
\end{tabular}
\hfill
\begin{tabular}{c||c}
 \multicolumn{2}{c}{Derived} \\
 \hline
 Variable & Grouping \\
 \hline
\hline
$\sum (p_T)$ &  \multirow{10}{*}{Momentum}  \\
$\slashed{E}_T$ &  \\
$H_T$ &  \\
$S_T$ &  \\
$p_T^{\ell_1}/p_T^{j_1}$  &  \\
$p_T^{\ell_1}/H_T$ &  \\
$S_T/p_T^{j_1}$ &  \\
$S_T/H_T$ &   \\
$p_T^{\ell_2}/p_T^{j_1}$  &  \\
$p_T^{\ell_2}/H_T$ &  \\
\cline{0-1}
$p_T^{\ell_1}/p_T^{\ell_2}$ &D-lepton: momentum  \\
\cline{0-1}
$|\Delta \eta_{\ell_1, \ell_2}|$ &  \multirow{2}{*}{Di-lepton: angular}  \\
$|\Delta \phi_{\ell_1, \ell_2}|$ &    \\
\
\end{tabular}
\caption{Variables used at the feature selection stage, where the primary variables denote low level variables taken directly from simulators, and derived variables are higher level and require manipulation. The final 5 derived variables are only considered if there are at least two leptons in the event ($n_\ell \geq 2$).}
\label{tab:features}
\end{table}

\subsubsection{Principal component analysis}
\begin{figure}
\includegraphics[width=0.46\textwidth]{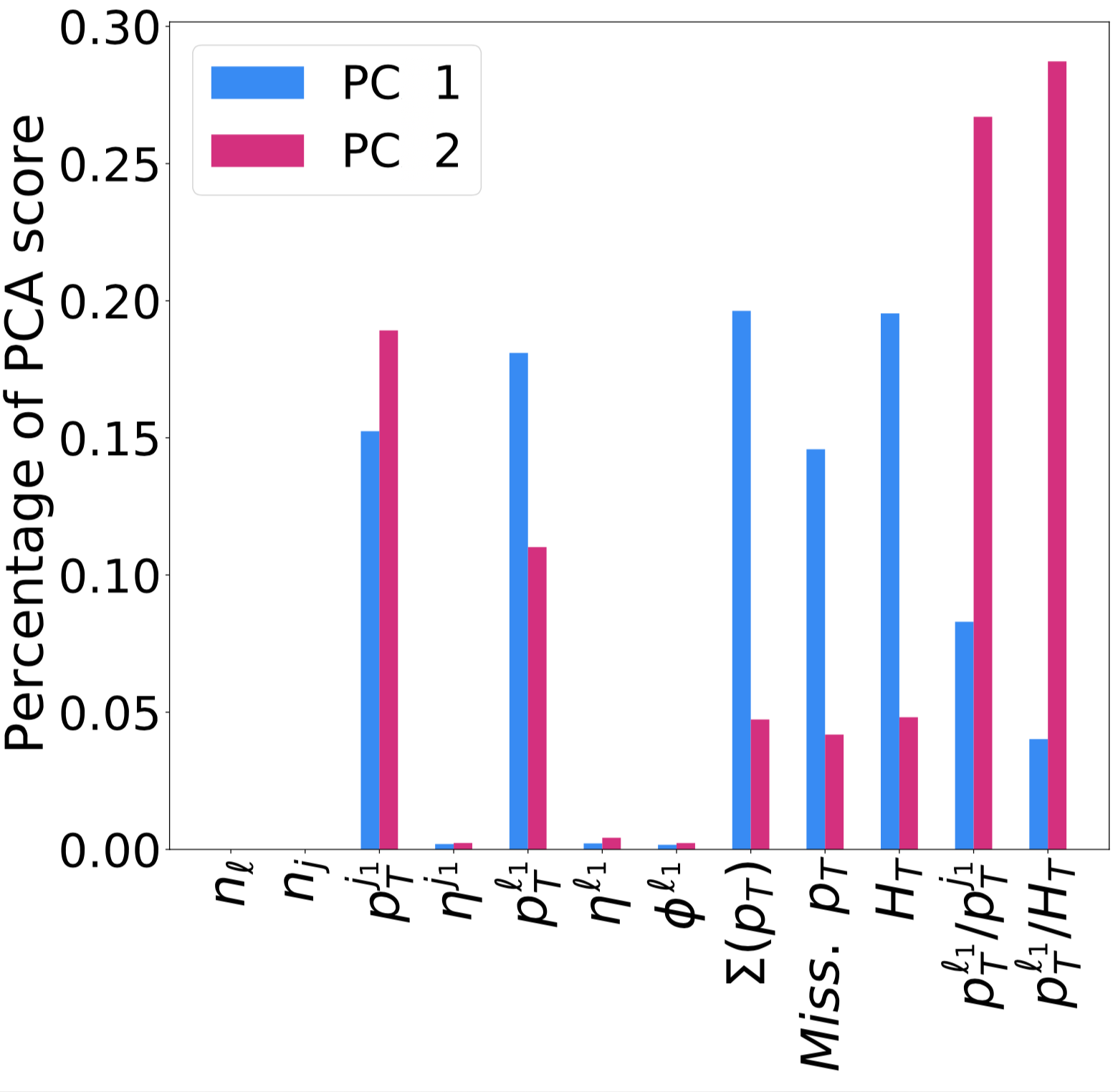}
\includegraphics[width=0.45\textwidth]{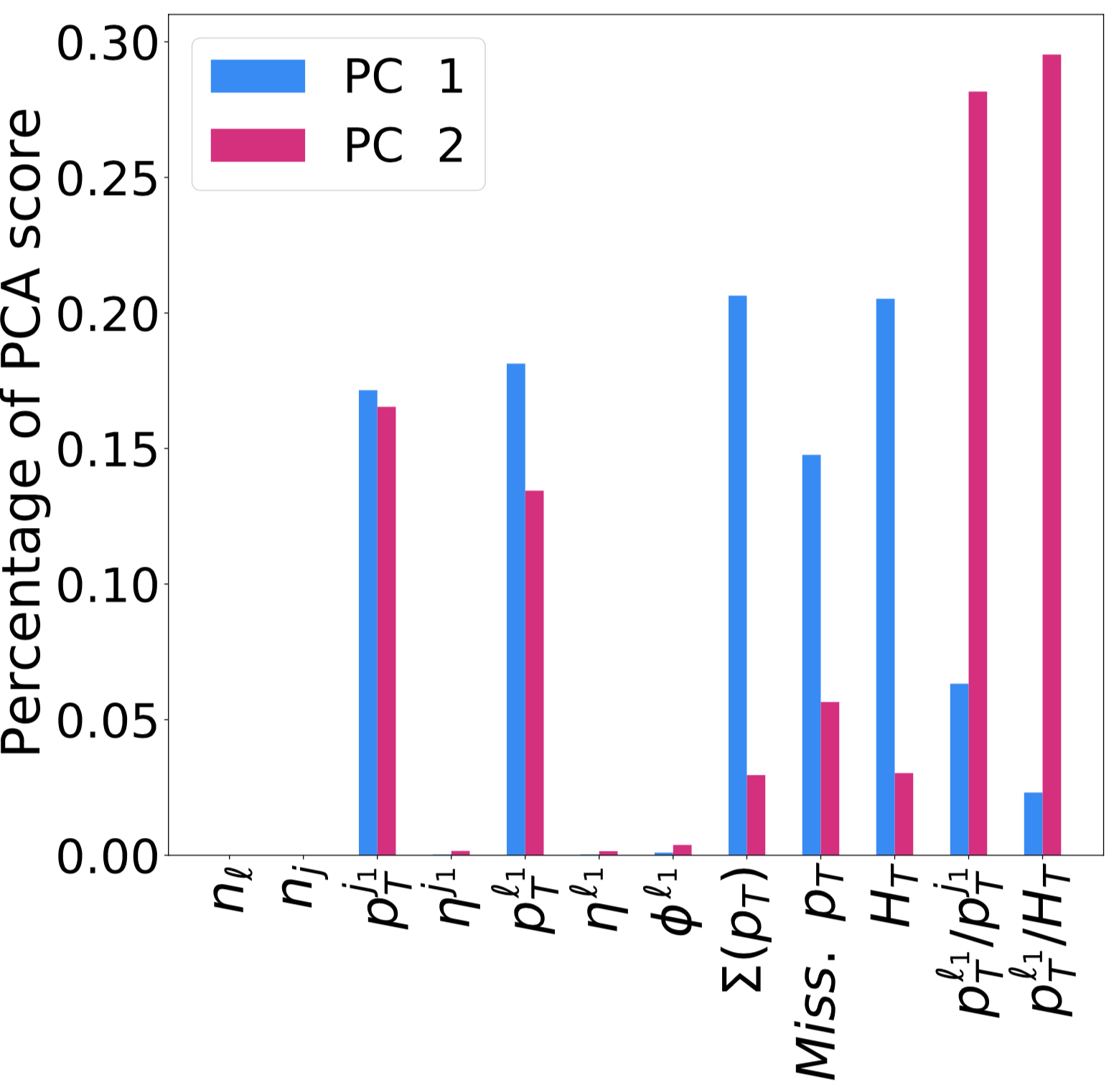}
\caption{PCA bar plot for the feature investigation step, including the features selected by the PCA analysis for two benchmark $(m_{\tilde{\mu}}, m_{\tilde{\chi}})$ cases: (210, 65) GeV on the left and (240, 135) GeV on the right. Similar features are identified for both benchmarks. }
\label{fig:PCAall}
\end{figure}

For an initial investigation of the features and their discriminating power, we utilise a PCA. This technique offers an analytical pre-processing step, enabling deeper insight into the available features by investigating their variance. The PCA is not (and should not be) used in isolation for feature selection, but rather as a confirmatory pre-processing step, used in conjunction with the Shapley values and permutation importance results below. The technique features an $m\times n$ matrix $A_n$, which transforms a random vector $X \in \mathbb{R}^m$ with covariance matrix $\Sigma_X$ to a lower dimension random vector $Y \in \mathbb{R}^n$ with $n<m$~\cite{10.1145/1291233.1291297}. Crucially, the matrix $A_n$ retains the variation of the original space within the lower dimensionality space, selecting features with large variance. The columns of the matrix $A_n$ are the $q$ orthonormal eigenvectors corresponding to the sorted eigenvalues, or principal components, of the covariance matrix. The new features are constructed through the linear transformation 
\begin{displaymath}
Y = A^T_n X,
\end{displaymath}
and capture the majority of the variance in the original data. In figure~\ref{fig:PCAall}, we gradually increase the number of features considered, monitoring their contribution to the two principal components (PC). The use of two PCs is the typical exploration technique, and we have found that a third PC adds negligible additional information. The features shown in the plot are the dominant contributors to the principal components. The left plot features a \textit{separated} scenario, and the right plot features a \textit{neutral} scenario, but the percentages of the PCA score obtained for each of the two principal components in each case are very similar. In particular, the same variables are favoured highly by  each of the PCs, where for instance in both the split and neutral scenarios the $p_T$ ratios are selected by the second PC, and the $H_T$ variable by the first PC. From this we deduce that the variance of the features across the three mass splitting regimes are similar.\

While the PCA is not used here to employ the principal components in developing the model, it allows for investigation of the features which contribute to the components and assessment of whether the most important features are those one would naively expect. The PCA offers a way to visualise the variance within the data, but the dimensionality reduction that it offers should not be confused with feature selection. Features with higher variation are not necessarily more useful in a physics analyses than those with lower variation, but variance analysis of features in an uncorrelated format does offer insight into how each of the features within the dataset will contribute to the target concept being learnt.\

Following this investigation into the feature behaviour, we are satisfied that momentum-based features are expected to play a significant role, as would be expected from the physics case. It is also of interest that only two of the ratios of eq.~(\ref{eq:ratios}) appear to be selected by the PCA, suggesting that the machine learning algorithm may be able to perform without needing all of the constructed variables which were used in the cut-and-count. We are therefore motivated to continue with a feature importance investigation which will use supervised learning to select the most valuable features.

\subsubsection{Feature permutation importance and Shapley values}

To employ feature selection, we turn to Shapley values, employed using the SHapley Additive exPlanations (SHAP) package. A concept first applied to game theory in 1953, Shapley values measure the average marginal contribution of each feature in enhancing the classification accuracy of the model as part of a group of features. The computed Shapley value is defined via a value function $\nu$ of all features in a set $S$, where $S$ is a set of elements explicitly exclusing element $x_j$ as $S\in\{x_1,...,x_p\}\backslash\{x_j\}$.  In this way, a given feature $x_j$ is isolated from the set, and the marginal value of the model without the feature and including the feature is calculated. The Shapley value is then calculated as~\cite{GomezRamirez785519}
\begin{equation}
\Phi_j(\nu) = \sum_{S\in\{x_1,...,x_p\}\backslash\{x_j\}}\frac{|S|!(p-|S|-1)!}{p!}(\nu (S \cup {x_j})-\nu (S)),
\end{equation}
with $p$ being the number of features in the subset $S$, $x$ is the vector of features under investigation, and $\nu$ is the prediction for the feature values. The Shapley value assesses the marginal gain (or loss) in outcome due to the parameter $x_j$ by calculating the value function with only the set $S$ as well as the set $S$ including $x_j$. In essence, the Shapley value of a feature is its contribution to the outcome, weighted and summed over all possible contributions. It can be employed in feature importance studies in both regression and classification tasks.\

\begin{figure}

\includegraphics[width=0.4\textwidth]{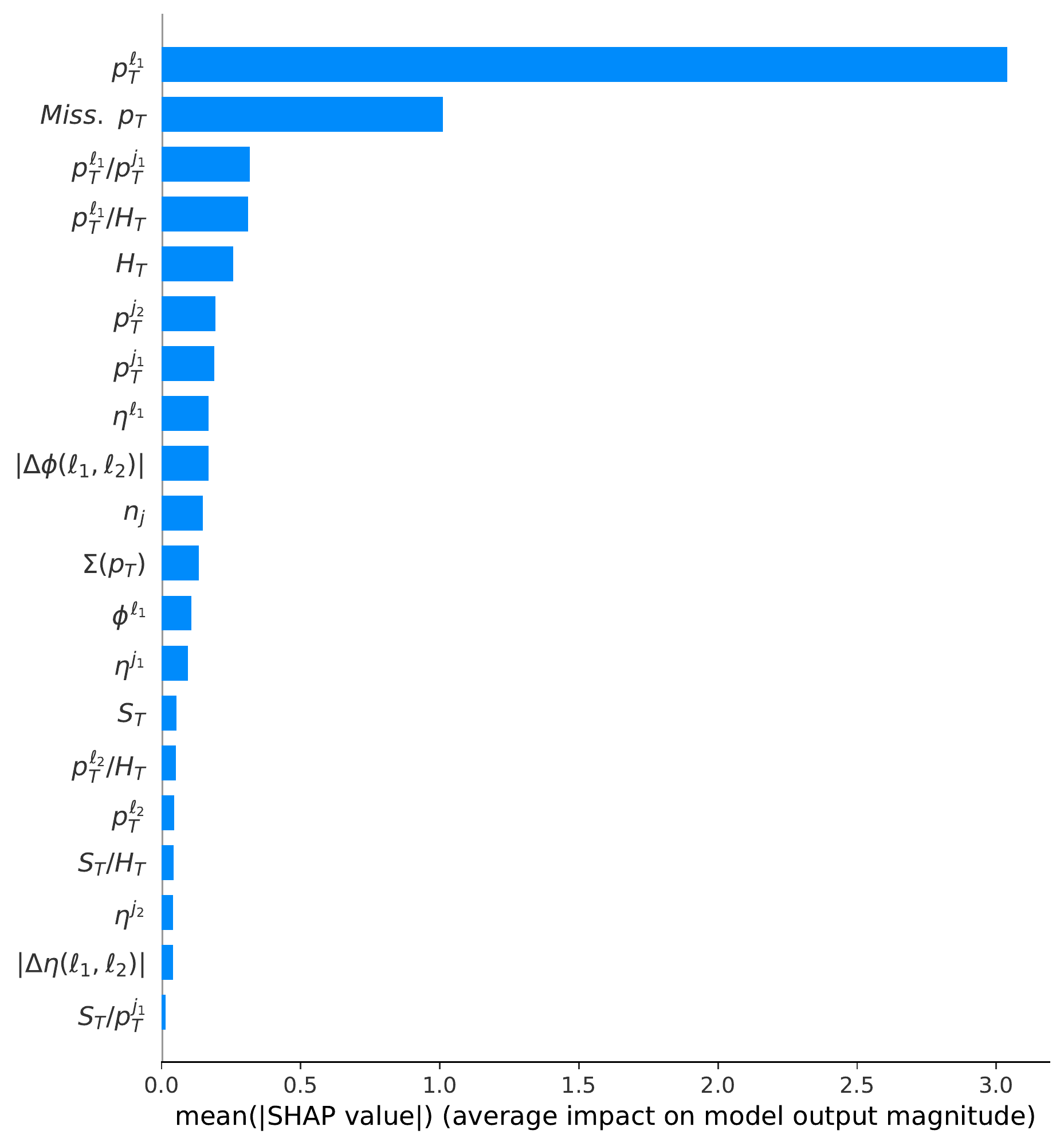}
\hfill
\includegraphics[width=0.4\textwidth]{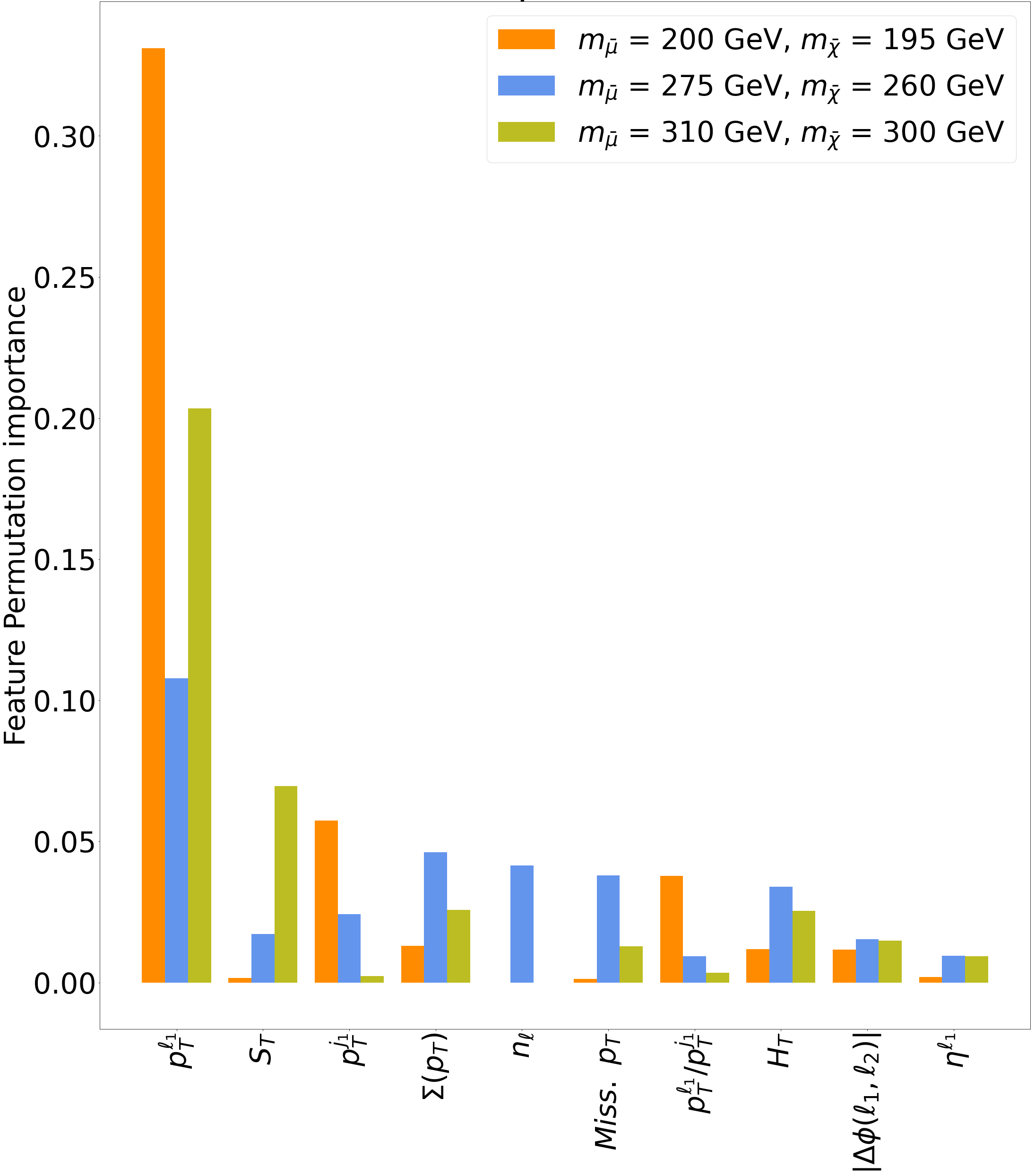}
\hfill
\includegraphics[width=0.4\textwidth]{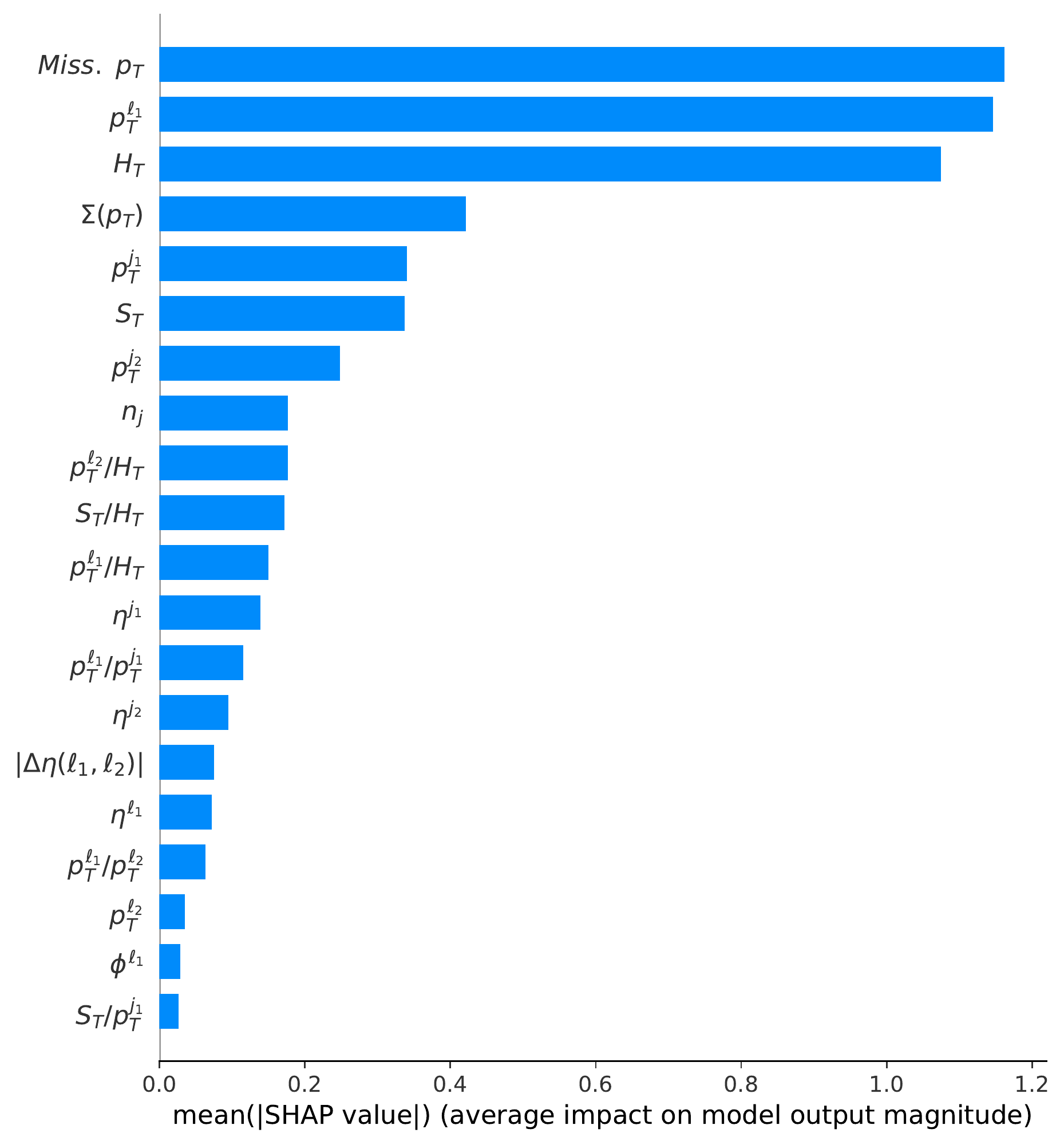}
\hfill
\includegraphics[width=0.4\textwidth]{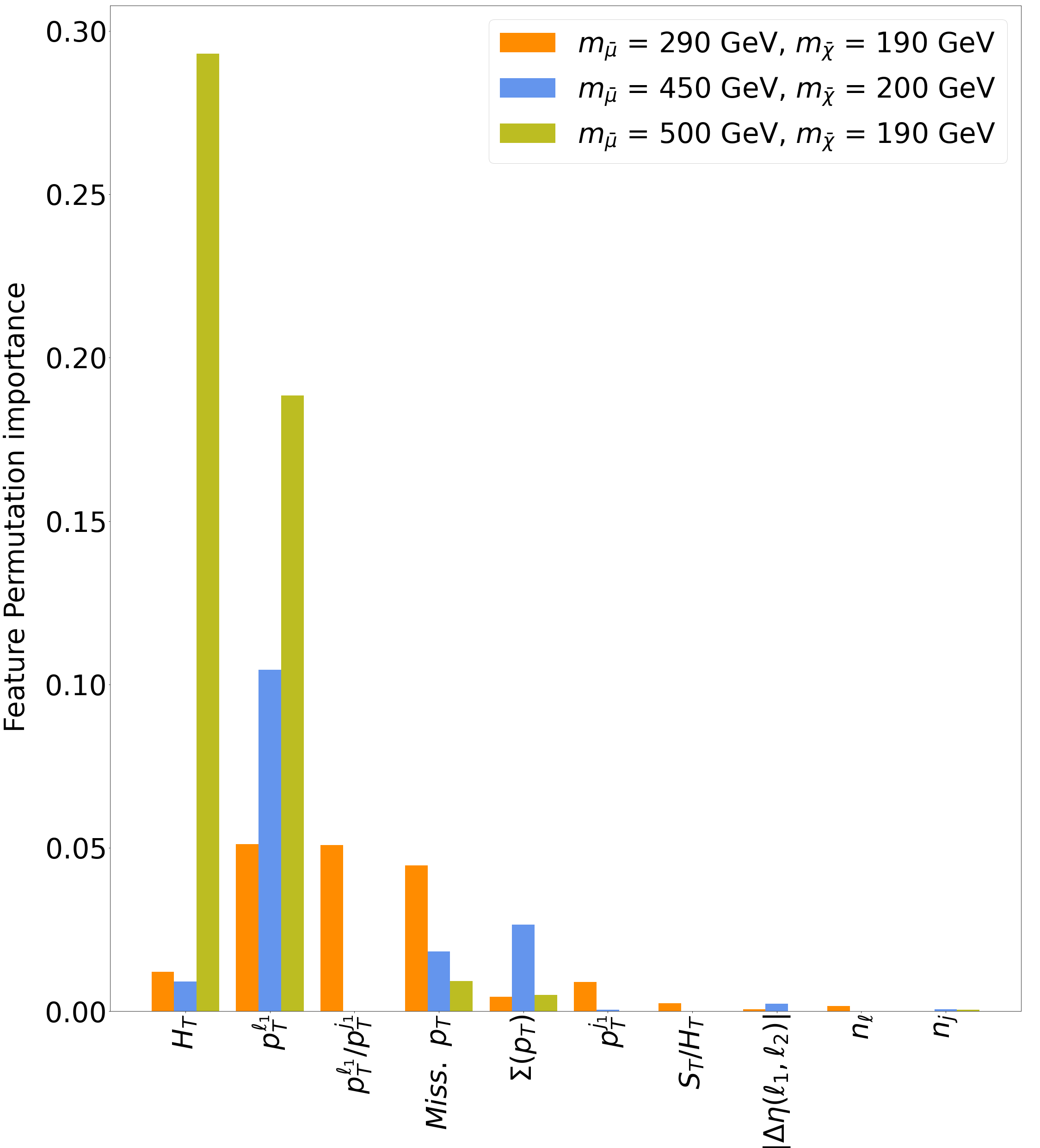}
\hfill
\includegraphics[width=0.4\textwidth]{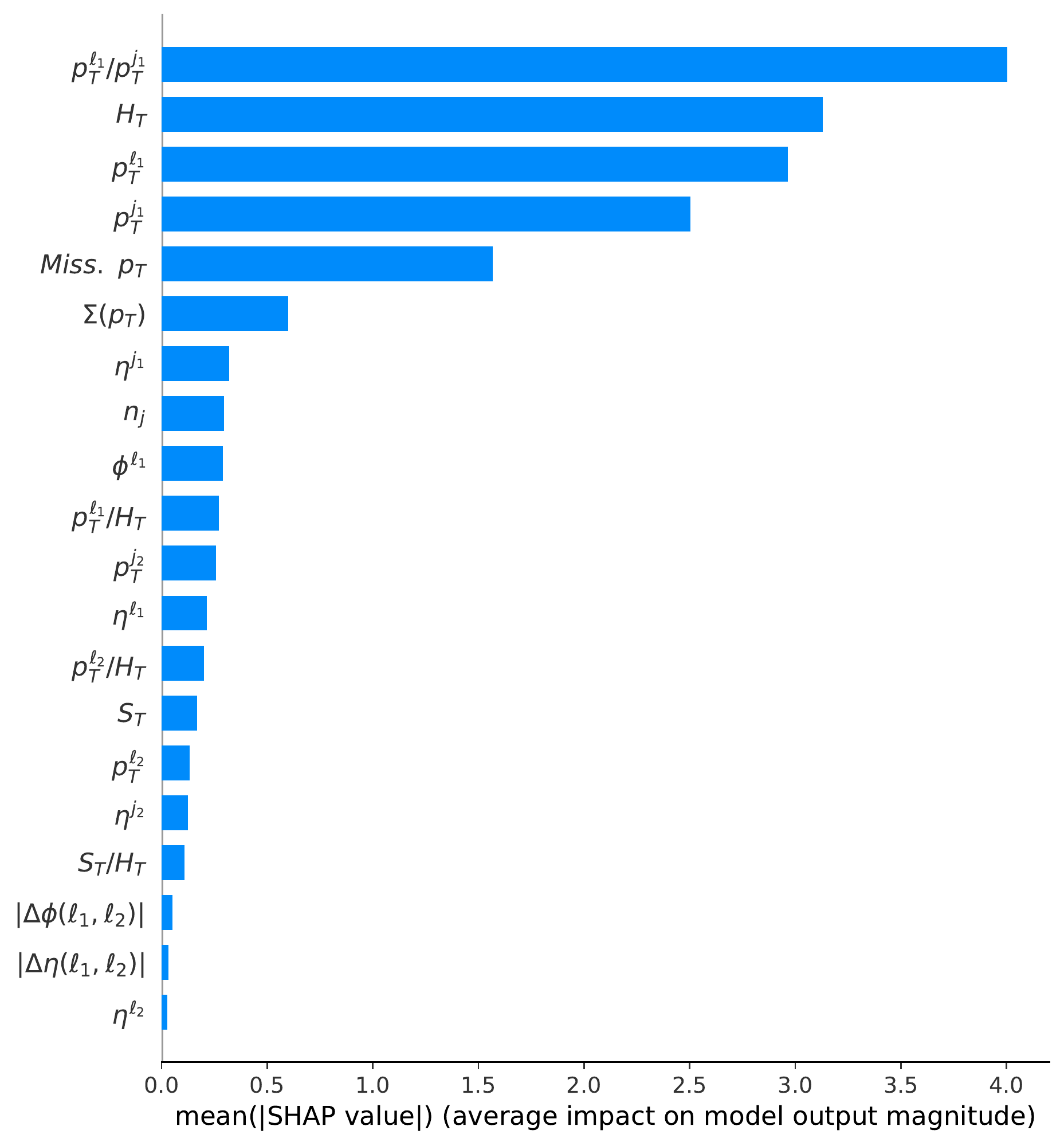}
\hfill
\includegraphics[width=0.4\textwidth]{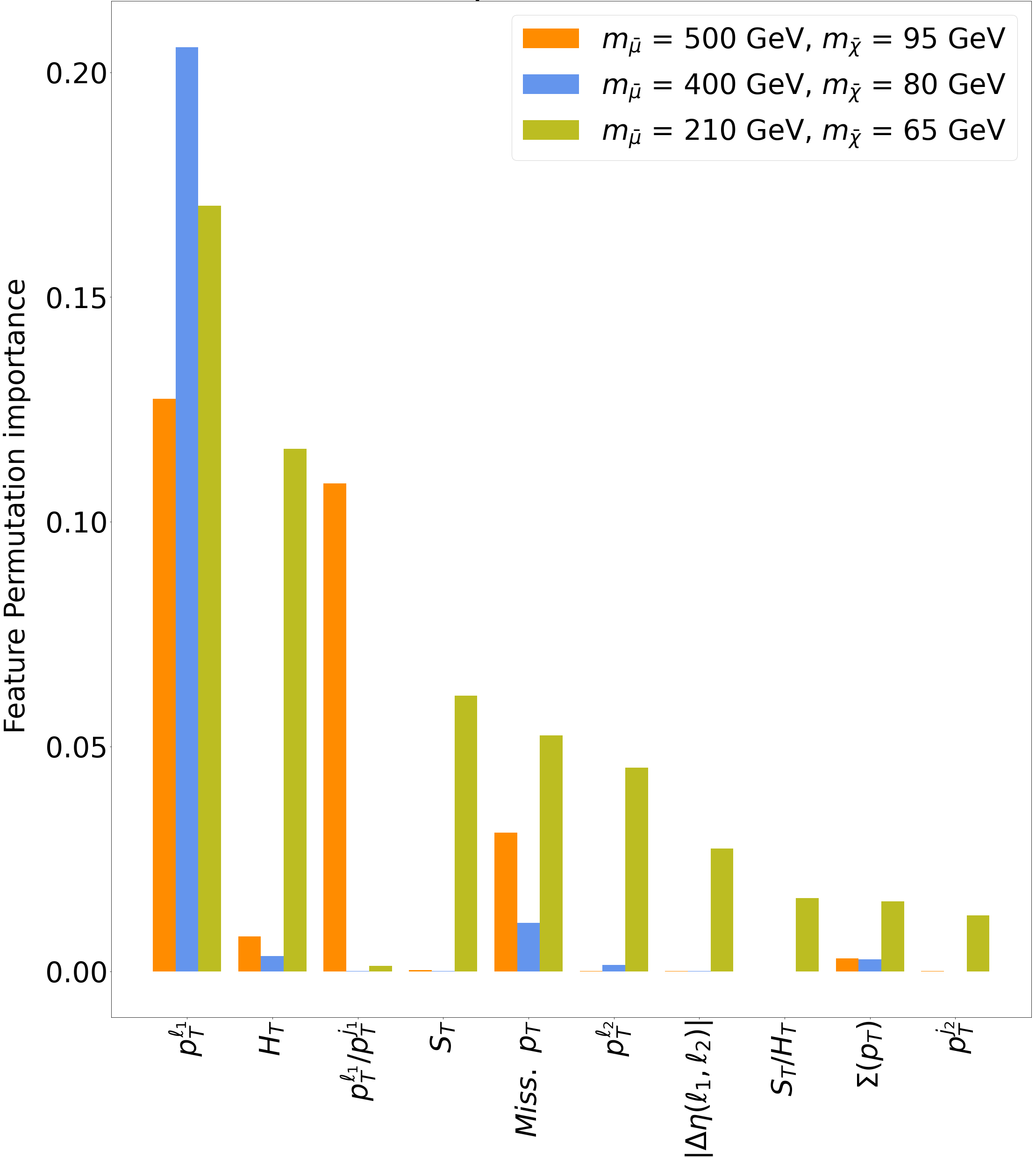}
\hfill
\caption{Sample Shapley feature importance plots (left) and permutation importance plots (right) for \textit{compressed}, \textit{neutral}, and \textit{separated} regimes. The Shapley plots are for $(m_{\tilde{\mu}},m_{\tilde{\chi}})~=~(200,195)$ GeV (top), $(m_{\tilde{\mu}},m_{\tilde{\chi}})~=~(290,190)$ GeV (middle) and $(m_{\tilde{\mu}},m_{\tilde{\chi}})~=~(400,80)$ GeV (bottom), while the permutation plots are divided into \textit{compressed}, \textit{neutral} and \textit{separated} mass ratios from top to bottom.}
\label{fig:pcasb}
\end{figure}

 Selecting only the most useful features can be crucial to avoiding overfitting in a machine learning task, as algorithms with too many input features can be too complex. In figure~\ref{fig:pcasb} (left), we plot the Shapley values for three benchmark values, including a \textit{compressed}, \textit{neutral} and \textit{split} case. The figure indicates the most important features, denoted by the length of the blue bars, in each scenario. It is evident that the most important feature in each case is unique; that is, different behaviour is observed across the mass ratios. In all cases, the five most important features are momentum-based variables, which may be expected from those variables being selected by the PCA in the previous section. Notably, many of the ratios presented in eq.~(\ref{eq:ratios}) feature in the Shapley selection, suggesting that they are important to the machine learning optimisation. Given the similarities between the PCA and Shapley investigation results, we are motivated to consider the final feature selection technique, which employs a permutation algorithm.

Decision trees rank and use features in order of selection power. In employing any feature selection, it is important to take the results with a grain of salt, and we should be careful how we interpret ranking. We consider the scenario where feature A and feature B are very similar, but B is always slightly less important than feature A, here feature B will always be ranked very low (as it does not add to the selection already made by feature A). However, if we then remove feature A, feature B will suddenly dramatically increase in importance. For this reason it is important to understand the physics variables employed in an analysis and their possible correlations, as well as carefully permuting through a feature importance investigation, which is employed in figure~\ref{fig:pcasb} (right). In this final feature investigation technique, the function uses a permutation to randomly shuffle features in a dataset, each time calculating the model score. Features are ranked by importance depending on the drop in model score when they are removed. In the figure, only the most important features are plotted, and they are arranged by importance.\

As can be seen in figure~\ref{fig:pcasb} (right), the features identified as being important across the spectrum of benchmarks are the
$p_T$ of the leading jet, 
 the $p_T$ of the leading lepton, 
the total transverse momentum $\sum p_T$, 
the missing $p_T$, 
the hadronic activity $H_T$, 
the $p_T$ ratio of the leading lepton over the leading jet, 
and the $p_T^{l_1}/H_T$ ratio. 
Less common is the pseudorapidity of the subleading jet, $\eta^{j_2}$, and the number of jets, $n_{jets}$. It is interesting to note that the features selected by the permutation importance algorithm, which come into play across all benchmarks, are all momentum-based variables, even though there are differing kinematics across the \textit{compressed}, \textit{neutral}, and \textit{separated} regimes. It is also clear that the \textit{compressed} regime places decidedly more relative importance on a sole variable, $p_T^{\ell_1}$, whereas \textit{neutral} and \textit{separated} regimes appear to be more balanced.\

In comparing the features selected as important by each algorithm, some similarities are evident; the lepton momenta and $H_T$ variables are always ranked highly, as is the missing transverse momentum. Interestingly, the angular variables $\phi$ are picked by the algorithm employing Shapley variables, but not by the permutation importance. Despite some differences, which are to be expected, there is clear agreement between all three methods in ranking momentum-based variables very importance. Additionally, it is notable that most of the ratio variables of eq.~(\ref{eq:ratios}) are not highly ranked by any of the methods, where it appears that using only a selection of the ratios is sufficient for the model.

\subsection{Tree structure}
\begin{figure}
\centering
\includegraphics[width=0.52\textwidth]{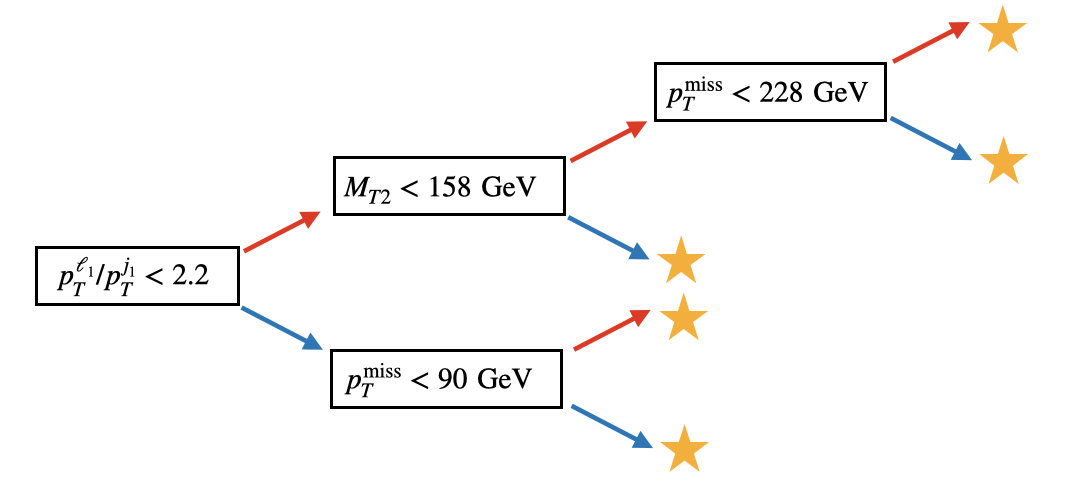}
\includegraphics[width=0.52\textwidth]{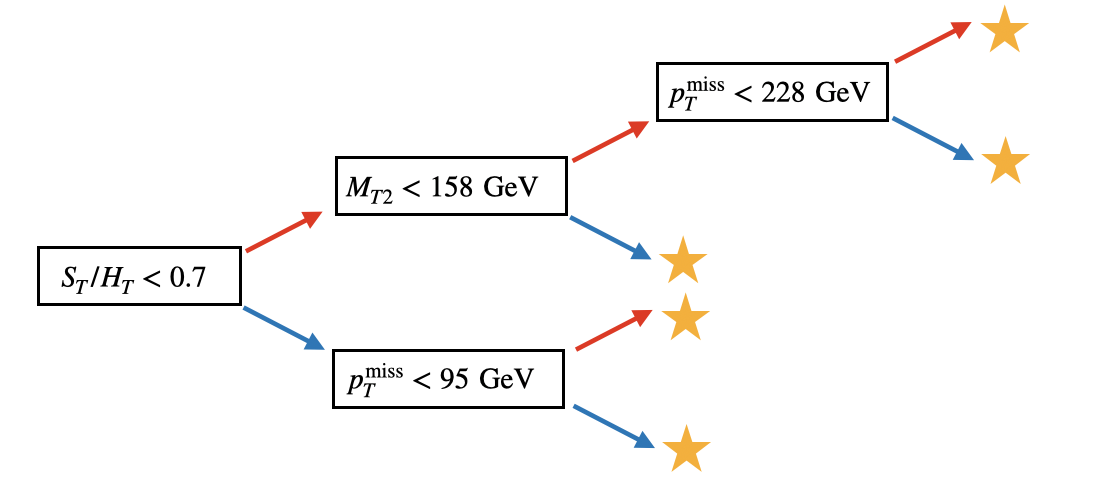}
\includegraphics[width=0.52\textwidth]{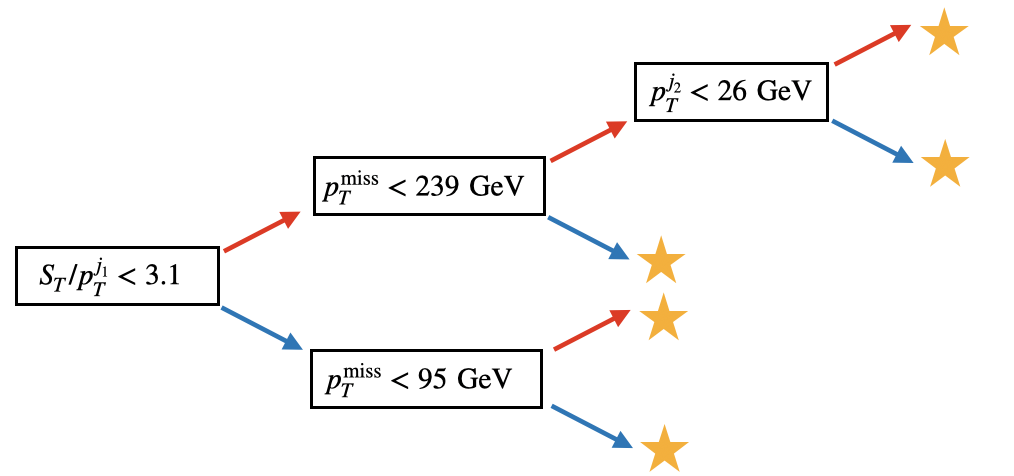}
\caption{The first, second, and tenth (top to bottom) trees built in the classifier trained on the benchmark $(m_{\tilde{\mu}}, m_{\tilde{\chi}}) = (300,100)~{\rm GeV}$. The red arrows indicate that the selection in the preceding box is satisfied, and the blue arrows indicate that either the event does not pass the selection or that the data is missing. Yellow stars indicate a leaf is reached. This model has a maximum depth of three.}
\label{fig:treestruc}
\end{figure}

In order to compare to the features selected as `important' by the Shapley values and feature permutation selections, we may examine the structure of the trees produced when a given model is trained. As an example, the decision tree trained for the $(m_{\tilde{\mu}}, m_{\tilde{\chi}}) = (300, 100)~{\rm GeV}$ scenario is presented in figure~\ref{fig:treestruc}. In this model, the maximum depth of any given tree in the model is 3, which was chosen by the hyperparameter selection outlined in the following section. The model proceeds by constructing a first tree (the top plot in the figure), and saving the events which are not correctly classified. It then constructs a second tree in an attempt to classify the previously incorrectly classified events, visible in the second row of figure~\ref{fig:treestruc}. Small changes are evident between the two trees, with the introduction of the $S_T/H_T$ variable being notable, where the value chosen for the cut on $S_T/H_T$ by the machine learning model in this instance also differentiates signal from background in Ref.~\cite{Fuks:2019iaj}. Finally, the tenth tree is presented on the bottom row, where the value of $S_T/p_T^{j_1}$ chosen to separate signal from background is much larger than the representative values of Ref.~\cite{Fuks:2019iaj}. This highlights the utility of the final trees in the machine learning model building sequence; they would not be expected to perform well on all data points, but are chosen to separate remaining signal events which may feature outlying behaviour. \

The visualisation of the tree largely supports the results of the feature importance plots in figure~\ref{fig:pcasb}, where it should be noted that later trees and other benchmarks make use of separate features not shown in this example. It should be noticed that several variables are used multiple times in the tree, highlighting the difference between machine learning tools and cut-and-count selection. Whereas in the latter a single selection is made on a feature for all events, the advantage of machine learning is that features may be reused in multiple ways. Variables which were ranked as important are used multiple times, where the value of the cut is modified each time. The use of multiple trees within a model again highlights the difference between the cut-and-count approach and that of a BDT algorithm.\

\subsection{Hyperparameter selection}

Finally, we turn to the hyperparameter selection for the {\sc XGBoost} model. The values of these parameters, which control the structure and functioning of the machine learning optimisation, directly affect the performance and predictive power of a machine learning toolkit~\cite{4e602701df284c70b06d479d4100c1d7}, and so should be carefully considered. It has long since been established that different hyperparameters work for different datasets~\cite{Kohavi95automaticparameter}, and also that the tuning of such hyperparameters for a given dataset improves the performance of a given machine learning algorithm with respect to its default parameters~\cite{4e602701df284c70b06d479d4100c1d7, olson2018datadriven, DBLP:journals/corr/abs-1208-3719}. Manual optimisation of hyperparameters is possible, but automatic methods are more easily repeated and generalisable~\cite{14f00e7a0861477a81f65b5c51f660f4}, and also can produce better results by covering larger parameter spaces. Hyperparameters also often depend on each other, making tuning them independently unreasonable~\cite{4e602701df284c70b06d479d4100c1d7}. Tuning may be carried out via a grid search~\cite{10.1007/978-3-642-45111-9_21}, where the dimensions of the space of hyperparameters is reduced to a finite set of values and evaluated, or the aptly named random search method~\cite{Andradottir2015}, where arbitrary configurations in hyperparameter space are tested, usually for a set number of iterations. Additional optimisation algorithms have also been studied~\cite{4e602701df284c70b06d479d4100c1d7}. It is not possible to predict which hyperparameters will best suit a given model before they are optimised, and one set of parameters can not be widely applied across datasets, although a suspected range of validity can be provided, and is discussed below. The hyperparameters should be optimised using an automated method, where assumptions about the data can be introduced via the task (logistic or linear regression, or classification) and via bounds on the appropriate hyperparameters.\

The hyperparameters can be divided into tree-specific parameters affecting each tree within the model, and boosting parameters which control the functionality of the model. The boosting-specific parameters include learning rate, which controls the step size of the gradient boosting,  and the number of estimators ($n_{\text{trees}}$). The two are correlated, as higher numbers of trees could lead to overfitting which is manageable by modifying the learning rate. The full list of hyperparameters considered are given in table~\ref{tab:hyperp}, and also includes the maximum depth of any individual tree, and the minimum child weight, which controls the conditions for node splitting. The $\gamma$ and $\alpha$ parameters are both regularisation parameters, used to add a penalty term to the loss function. Finally, the subsample parameter randomly selects a portion of the events to be fed to the model.

In this work, a random search has been implemented to scan over the hyperparameters. Random search algorithms initialise a starting hyperparameter configuration, which is extended at each subsequent iteration by randomly generated hyperparameters~\cite{4e602701df284c70b06d479d4100c1d7}. The process is expected to be more efficient than a grid search algorithm, which becomes computationally expensive for higher-dimensional problems. The optimal hyperparameters were obtained using a random search algorithm, where 4-fold cross-validation with a replacement technique was utilised to assess the performance of the model with each hyperparameter selection. Where relevant, the hyperparameters are trained with either the \textit{F-score} and the \textit{auc} metrics, depending on which metric is to be used in further training and testing.

A number of hyperparameters were considered in the scan, which control both the individual trees and the functioning of the model as a whole. The values for which a uniform distribution was chosen are the parameters which can take on all real values, as such, it is recommended that they be passed as a continuous distribution. Validation is performed with a replacement. The number of trees trained (number of boost rounds) was set to 200, with early stopping employed. The hyperparameter scan ranges and an example set of results in presented in table~\ref{tab:hyperp}. Some variation is visible across the benchmarks; values of 2, 3 and 4 are selected for maximum depth, while a learning rate of $0.1$ was consistently chosen. Consistency in hyperparameter choice across mass regimes was not visible, where the compressed regime benchmarks featured hyperparameter choices across the full scan range, and many different hyperparameters were chosen across benchmarks.  

\begin{table}
 \renewcommand\arraystretch{1.5}
 \setlength\tabcolsep{8pt}
\begin{tabular}{c|c|c|c}
Hyperparameter & Description & Range & Value \\
\hline
\hline
\multirow{2}{*}{Learning rate} & Magnitude of change between sequential  & \multirow{2}{*}{$[0.0001, 1]$} & \multirow{2}{*}{$0.1$} \\
& trees $F_m(x) = F_{m-1}(x) + \eta_{LR} h(x)$~\cite{FRIEDMAN2002367}& & \\
\hline
\multirow{2}{*}{Max. depth} & Number of nodes along longest & \multirow{2}{*}{$\{2,3,...,8\}$} & \multirow{2}{*}{$2$} \\
& path from first node to a leaf & & \\
\hline
\multirow{2}{*}{Min. child weight} & Lower bound on number of events in a & \multirow{2}{*}{$[1,5]$} & \multirow{2}{*}{$2.14$} \\
& node required to undergo further splitting & & \\
\hline
\multirow{2}{*}{The $\gamma$ parameter} & Regularisation parameter used to control & \multirow{2}{*}{$[0,5]$} & \multirow{2}{*}{$2.86$} \\
& overfitting (also called min\_split\_loss) & & \\
\hline
\multirow{2}{*}{The $\alpha$ parameter} & Controls L1 (Lasso) regularisation~\cite{Lasso1, lasso2} & \multirow{2}{*}{$[0.1, 1.0]$} & \multirow{2}{*}{$0.23$} \\
& using a penalty term for the loss function & & \\
\hline
\multirow{2}{*}{Subsample} & Random sampling of X\% of the dataset & \multirow{2}{*}{$ [0.01, 1.0]$} & \multirow{2}{*}{$0.43$} \\
& for training & & \\
\end{tabular}
\caption{Hyperparameters used by {\sc XGBoost}. The hyperparameters are specific to the mass benchmark on which the {\sc XGBoost} toolkit is to be applied. The value indicated as an example corresponds to the $(200,60)$ benchmark as an example.}
\label{tab:hyperp}
\end{table}
Once the optimal hyperparameters for each model have been chosen, we are ready to train the machine learning tool on each mass benchmark to obtain the models for use in testing. In the following, each {\sc XGBoost} model is trained on a given benchmark, following which the trained model is applied to test data.

\section{Boosted Decision Tree application and results}
\label{sec:results}
Following the preparatory steps outlined in the previous section, we are now ready to train and test our model. In the following, we compare the performance of the \textit{auc}, \textit{F-score} and \textit{ams} metrics on our model performance, outlining the case where the widely-used \textit{auc} is less reliable than the $F$-score metric. The metrics are implemented within a loss function, and while a BDT is able to include multiple metrics in a given training, we choose instead to compare the individual performances of the \textit{F-score} and \textit{auc} metrics. This is in order to assess the standard way in which metrics are implemented in generic particle physics analyses of this nature, where traditionally only the \textit{auc} is employed.

Finally, we compare the results of the BDT algorithm to those of the cut-and-count, investigating the generalisability of the models and assessing the use of each metric. We compare the performance of the BDT algorithm under the  metrics of interest, considering as a reference the exclusion region indicated in Ref.~\cite{Fuks:2019iaj} as related to the use of the dynamic jet vetoes. 

\subsection{Feature utility}
\begin{figure}
\includegraphics[width=\textwidth]{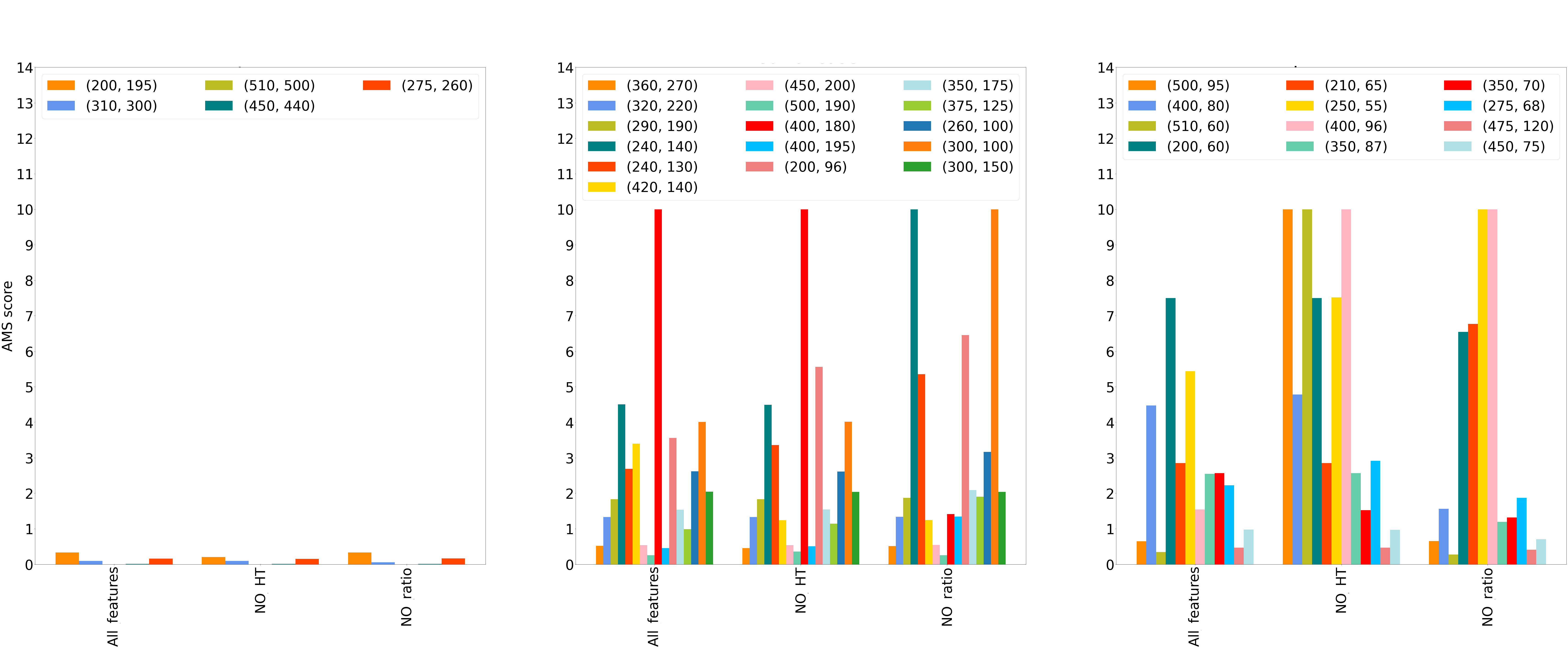}
\includegraphics[width=\textwidth]{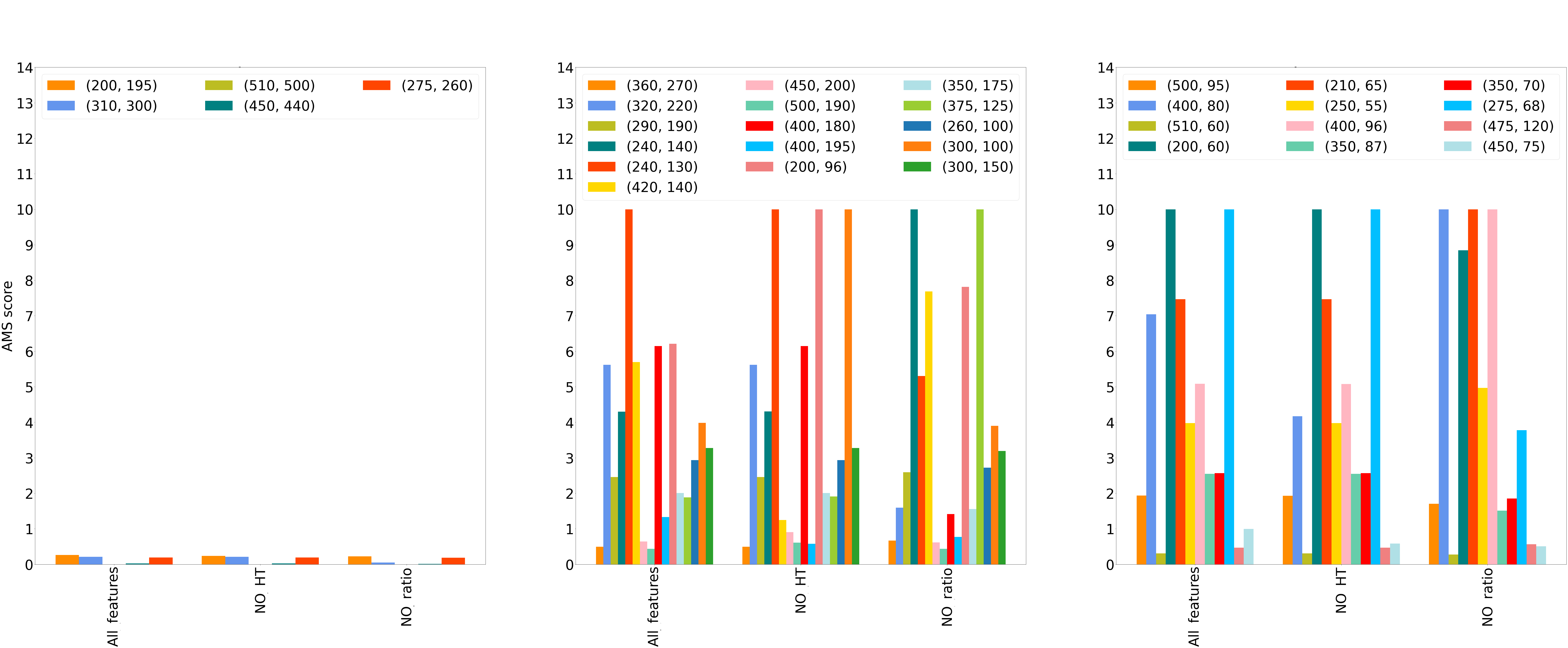}
\caption{The  top figures show the {\it ams} score of the model when using the {\it auc} metric (top) and the {\it F-1 score} (bottom) to train the model, with \textit{compressed}, \textit{neutral} and \textit{separated} scenarios from left to right. The masses specific to the benchmark are labelled ($m_{\tilde{\mu}}, m_{\tilde{\chi}}$).} 
\label{FIG:AMSSCoreComparisonPlots}
\end{figure}

The cut-and-count analysis of Ref.~\cite{Fuks:2019iaj} utilised sophisticated mass ratios which were chosen to provide a better discrimination between the signal and backgrounds. The utility of these ratios in a BDT algorithm were investigated first, by training and testing the BDTs at each benchmark with and without the ratios. We define three test cases; all ratios included, $H_T$ and $S_T$ values removed but ratios included, and ratios removed with $H_T$ and $S_T$ values included. We compare the use of the \textit{auc} and \textit{F-score} metrics on the training set, which is a subset of the full dataset. We compare the use of the \textit{F-score} and \textit{auc} metrics in each case (including all ratios, including only $H_T$ and $S_T$ and including no ratios or $H_T$ and $S_T$) by computing the \textit{ams} score of the trained models on the test data. The results are presented in figure~\ref{FIG:AMSSCoreComparisonPlots}, where \textit{compressed}, \textit{neutral}, and \textit{separated} benchmarks are presented for several feature cases. The figure is interesting as it reveals not only the behaviour of the benchmarks across the different feature spaces, but also the different behaviours of machine learning models within one mass regime. Where one may expect the BDT to be incomparable to other mass regimes due to the individual tuning of its hyperparameters in each case, we find instead that some degree of generalisability is evident. \

Figure~\ref{FIG:AMSSCoreComparisonPlots} reveals models performing better in general using the \textit{F-score} metric than the \textit{auc}, as well as different behaviour across the benchmarks. While in general better performance is achieved using the \textit{F-score} metric, the relative performances of each benchmark was inconsistent across the cases. We also note similar performance both when including the mass ratios and when removing them. In fact, it could be argued that better performance is visible when removing them. This suggests that, particularly when utilising the \textit{F-score}, the use of sophisticated derived features which were necessary in the cut-and-count analysis are not crucial to the functioning of the BDT model. This is encouraging, as it allows users to work with simpler features.\

\begin{figure}
\centering
\includegraphics[width=0.49\textwidth]{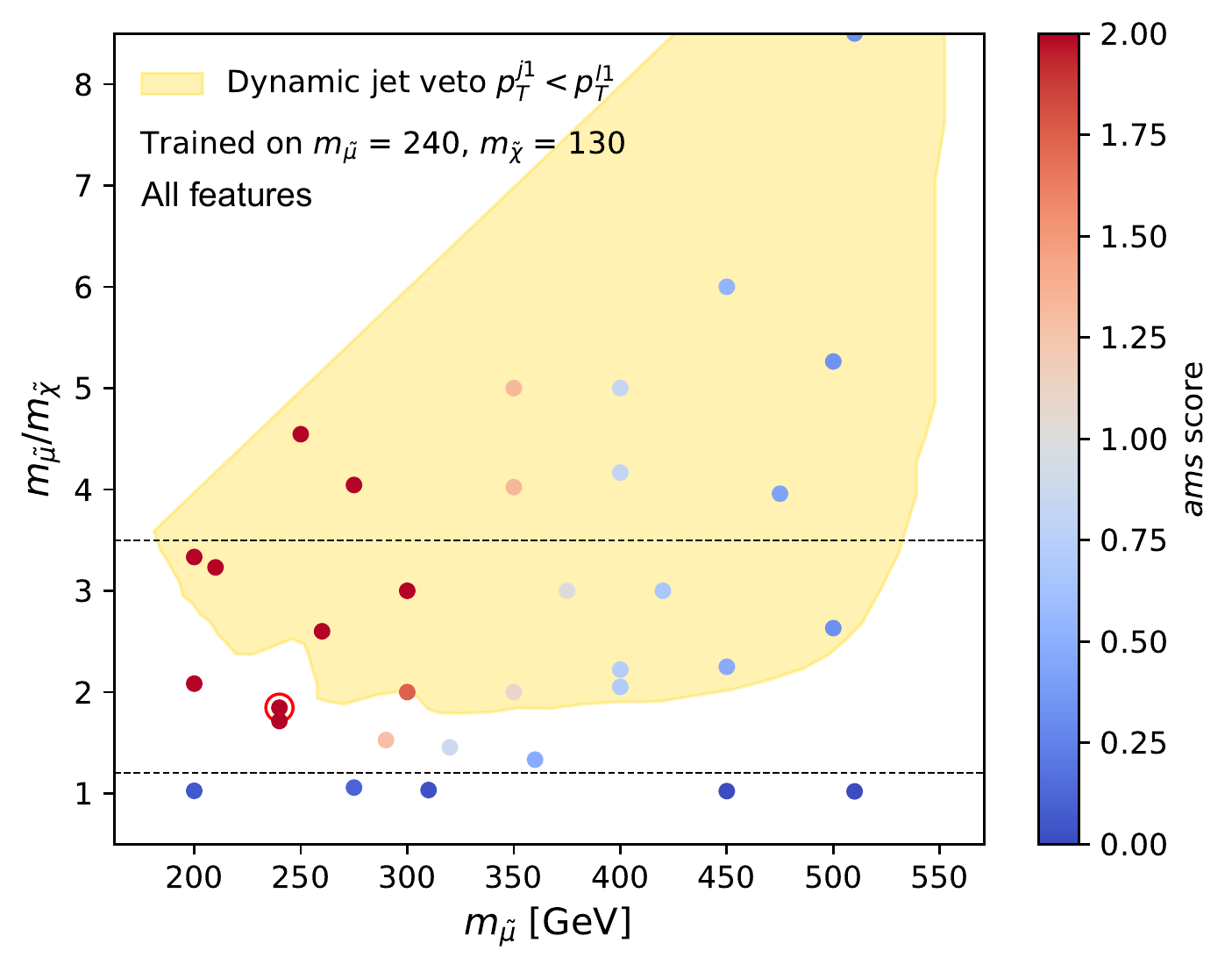}
\includegraphics[width=0.49\textwidth]{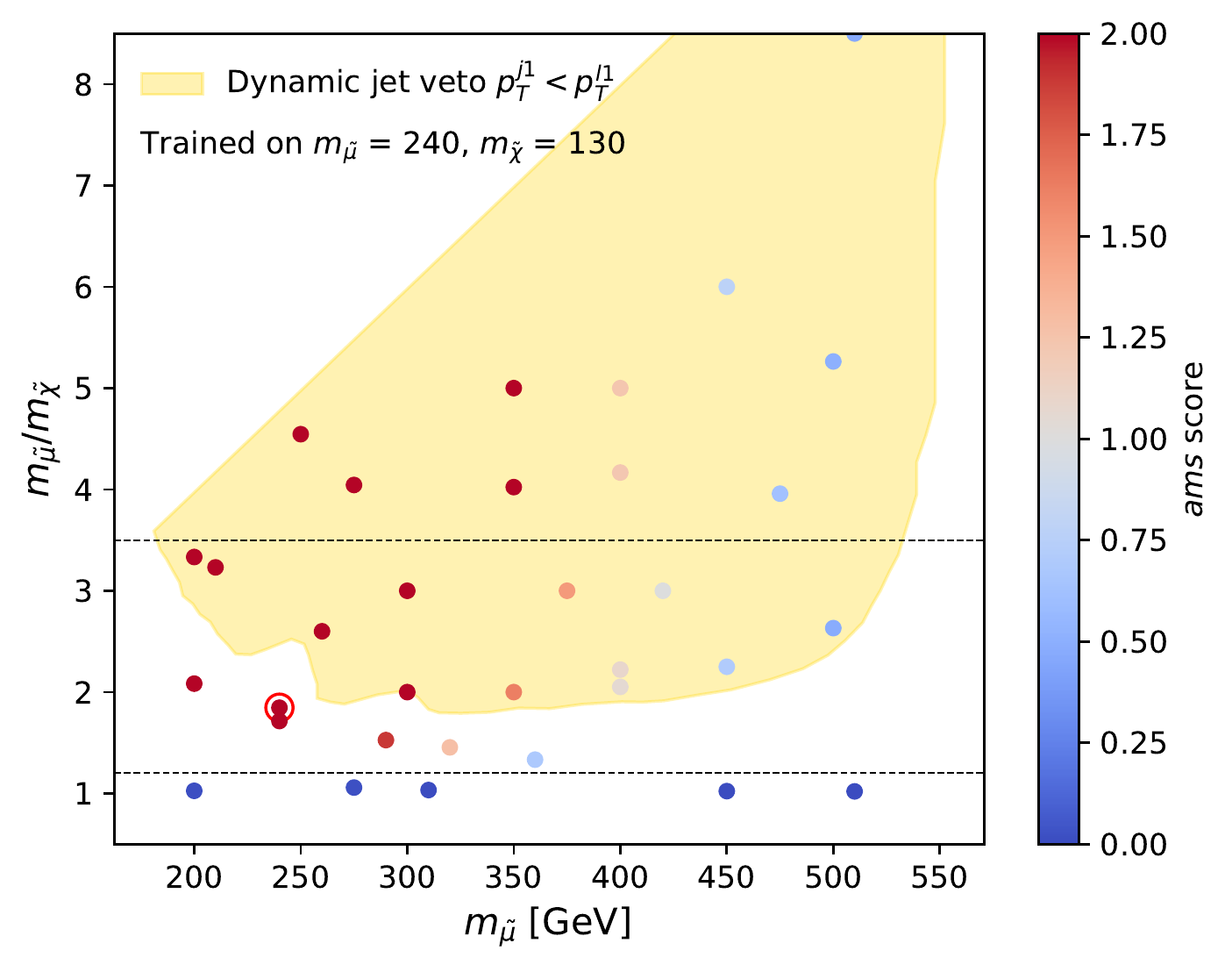}
\includegraphics[width=0.49\textwidth]{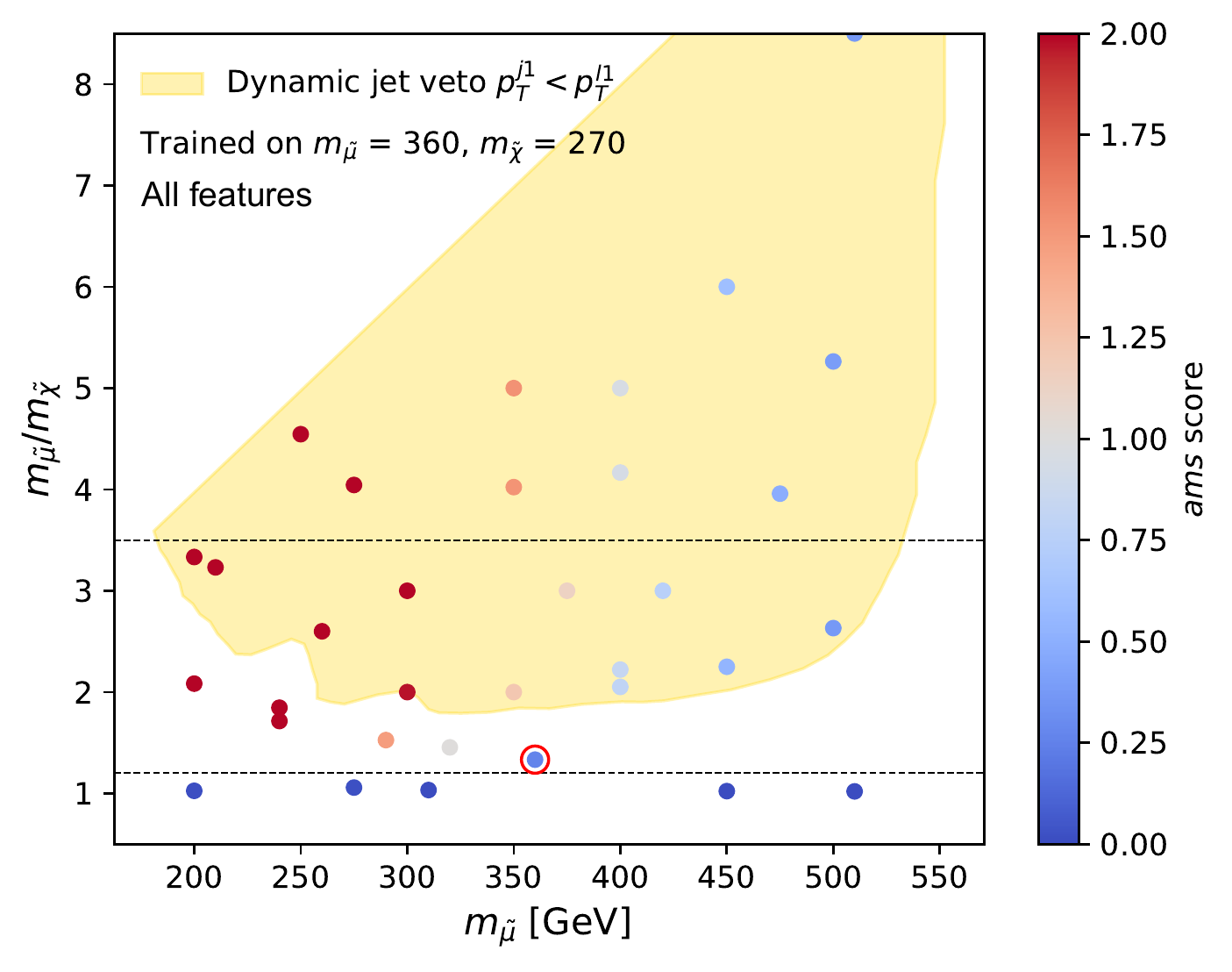}
\includegraphics[width=0.49\textwidth]{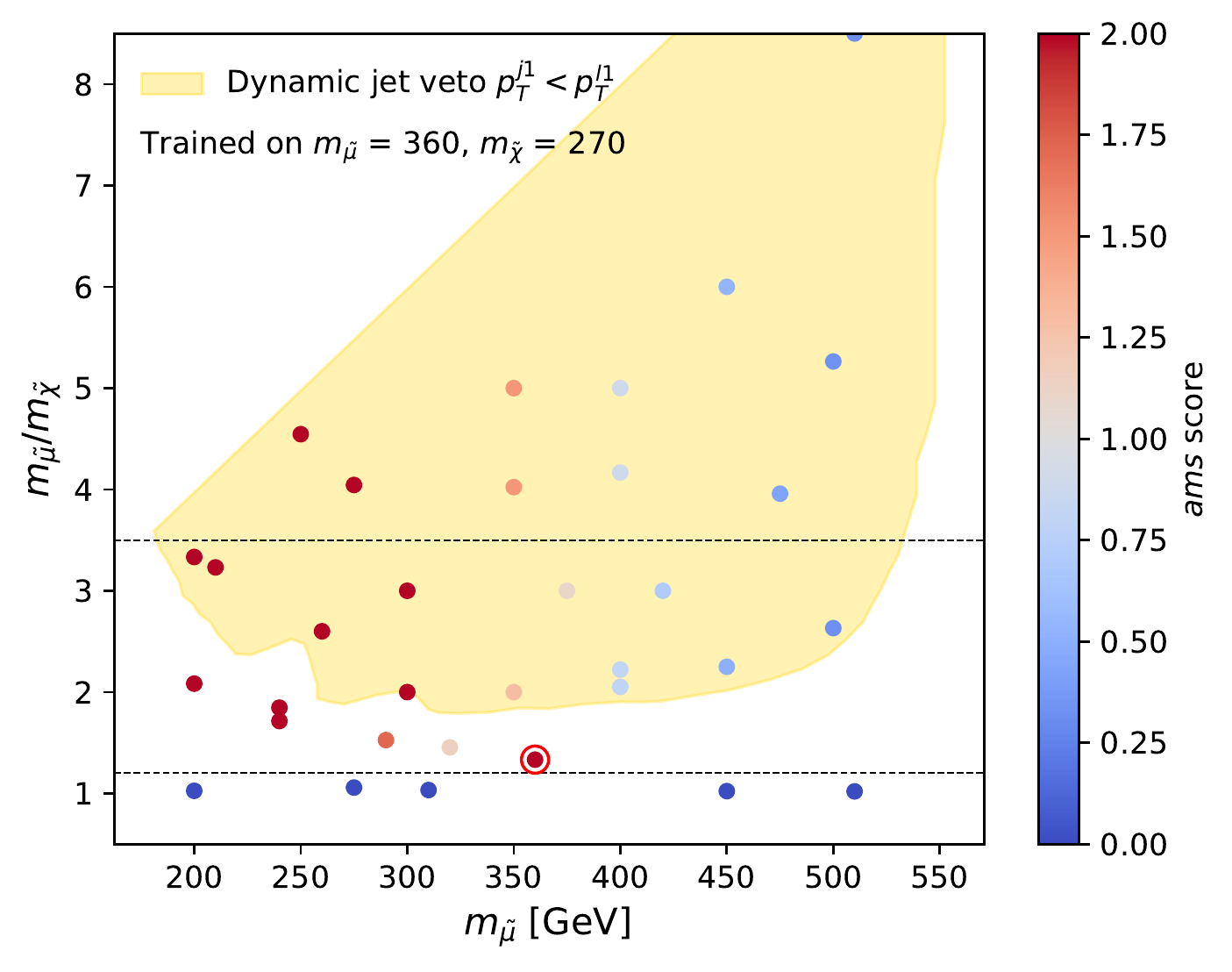}
\caption{Comparison of the `all features' case (left) with that of no ratios (right), showing better reach (top line) and better performance on the training benchmark (bottom line).}
\label{fig:exclfeaturecomp}
\end{figure}

To support this claim, we present a comparison of performance with and without the ratios in figure~\ref{fig:exclfeaturecomp}, where the benchmark on which the model was trained is encircled in red. The yellow region indicates the area where $2\sigma$ was achieved in the cut-and-count. The figure shows a comparison of two benchmarks, where the inclusion of all features is shown on the left, and removal of ratios on the right. In $(m_{\tilde{\mu}}, m_{\tilde{\chi}}) = (240,130)~{\rm GeV}$ we see good performance on the training benchmark in both cases, but the model is generalisable to more benchmarks when removing the ratios. In the case of the $(m_{\tilde{\mu}}, m_{\tilde{\chi}}) = (360,270)~{\rm GeV}$ benchmark, including all features does not lead to a good $ams$ score on the training benchmark, but this is improved by removing the ratios. We emphasise this point, noting the following.

\noindent\fbox{\parbox{\textwidth}{While the cut-and-count analysis relied on the use of dynamic jet veto ratios, the machine learning tool achieves equally, if not improved, performance without them.}}

The reason for the poorer performance while including the ratios could be one of complexity. It is possible, for example, that the inclusion of the ratios as well as the independent variables in the ratios is leading to a model which overfits or is too complex. The ratios may also translate less predictably to other mass cases, while the individual features are more consistent. In what follows, we therefore choose to omit the ratios when training and testing our model.

\subsection{Model generalisability}
\begin{figure}
\centering
\includegraphics[width=0.49\textwidth]{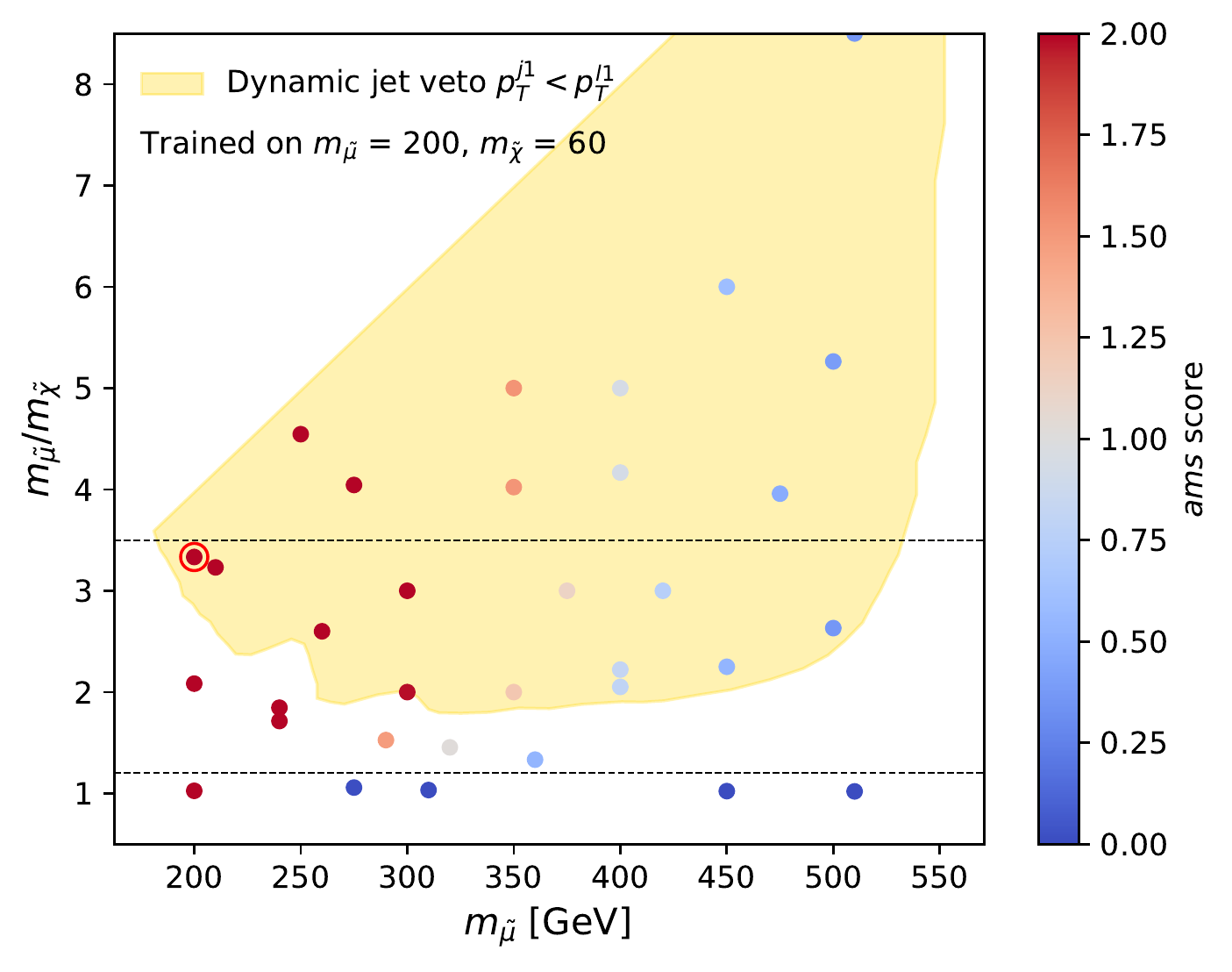}
\includegraphics[width=0.49\textwidth]{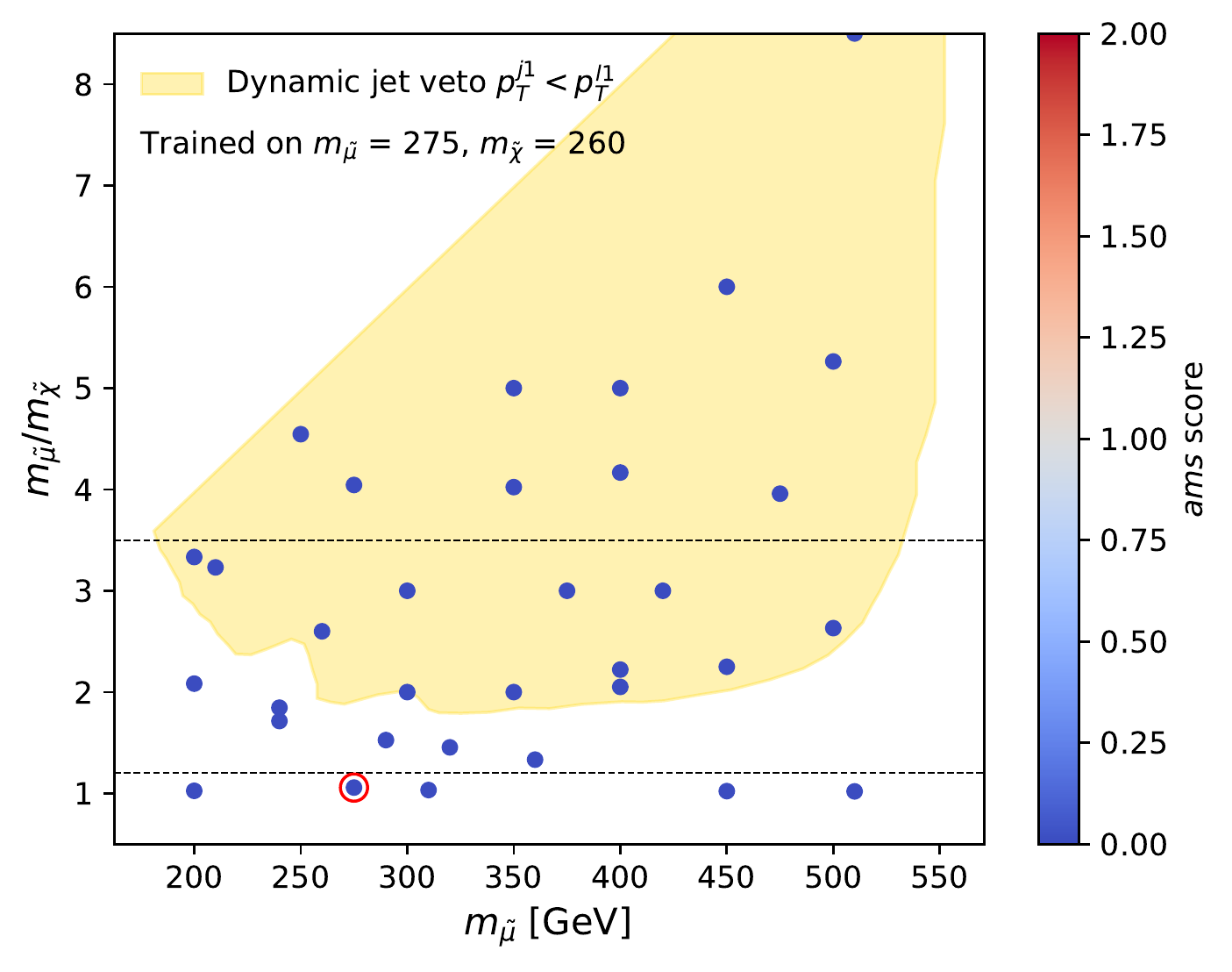}
\includegraphics[width=0.49\textwidth]{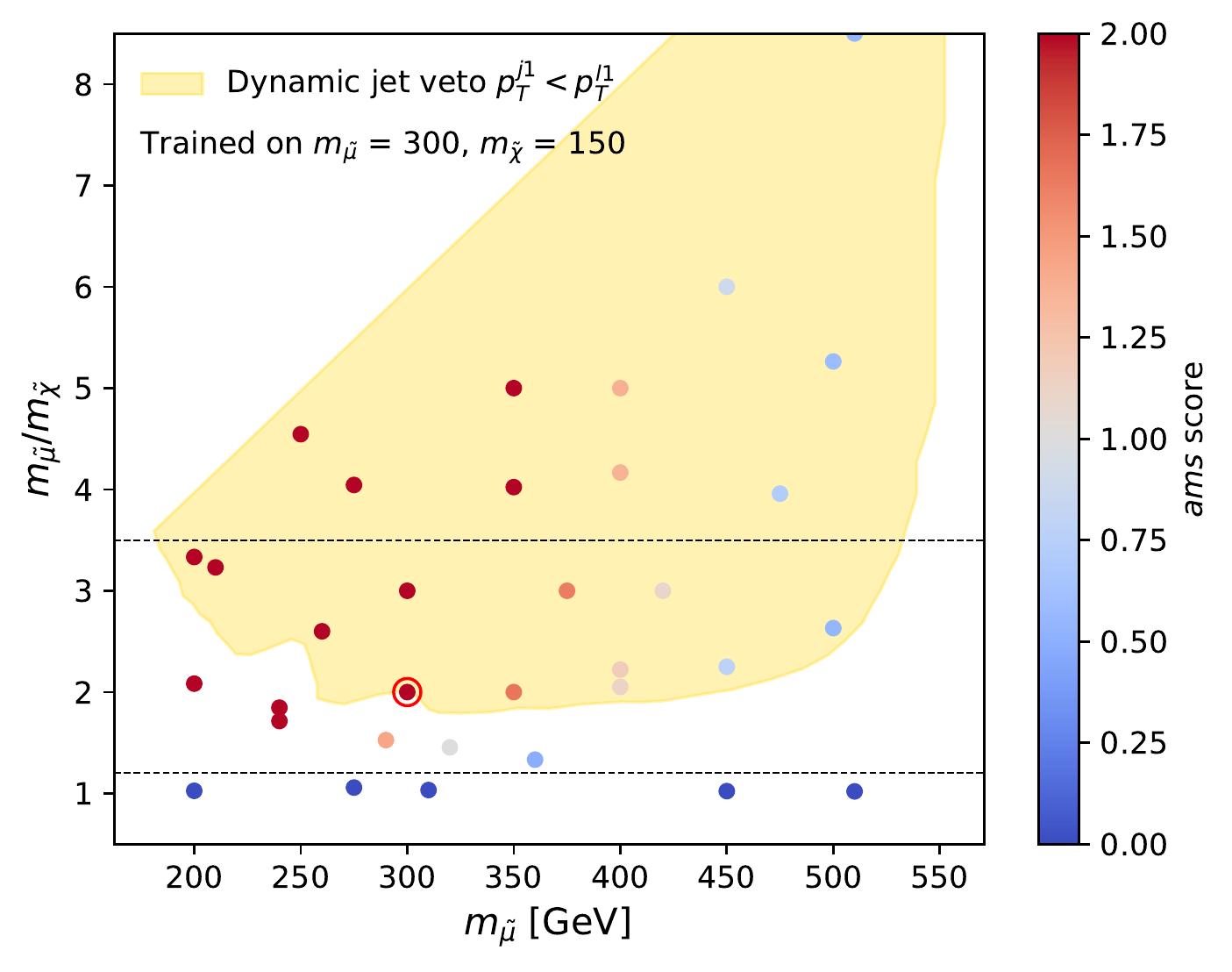}
\includegraphics[width=0.49\textwidth]{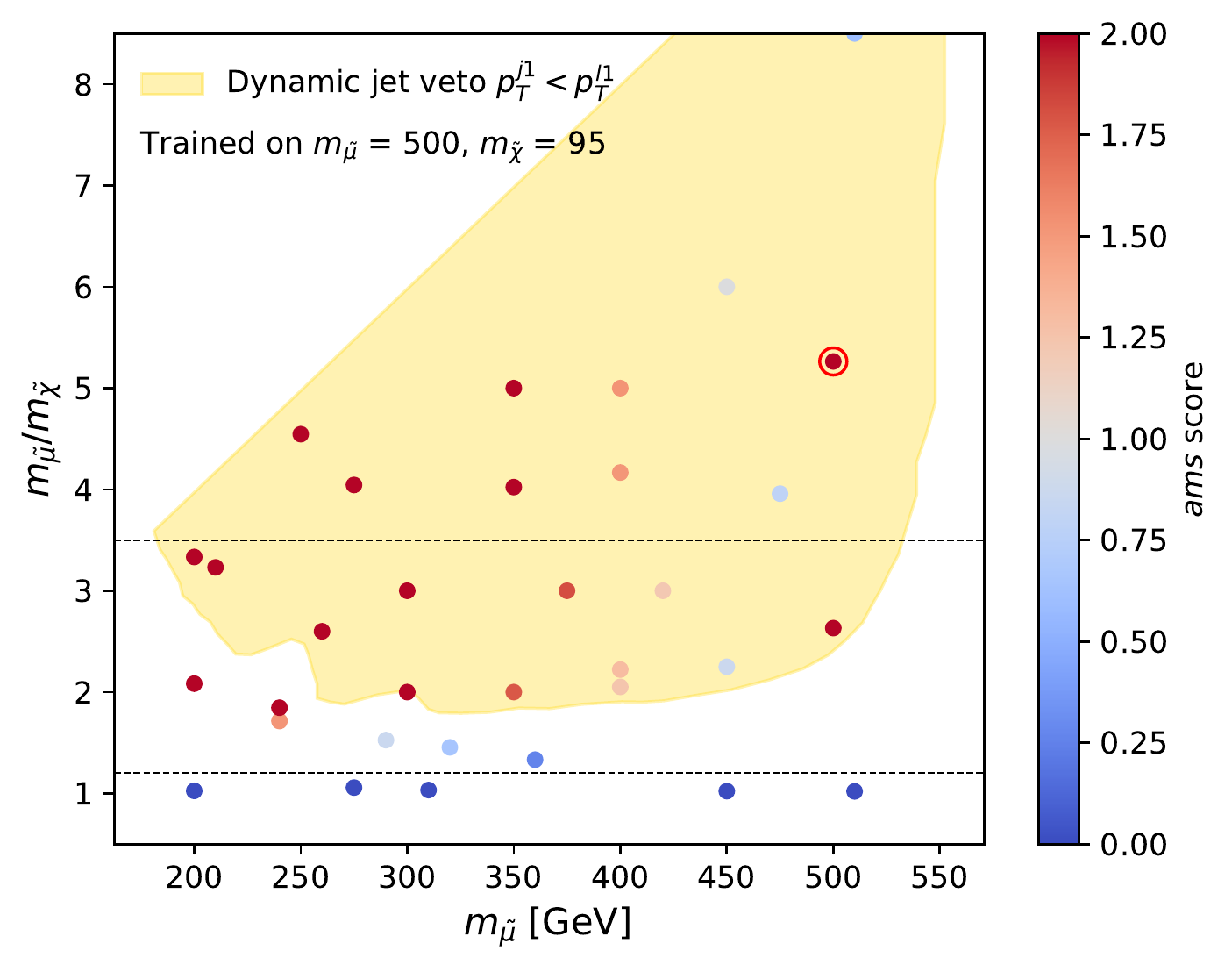}
\caption{Good performance of a model trained on a specific benchmark is presented, where the training mass point is circled in red. The shaded yellow region corresponds to the 95\% CL exclusion obtained using the dynamic jet veto in Ref.~\cite{Fuks:2019iaj}. The designation of the mass regimes in given by the dashed black horizontal lines. Points considered are randomly generated.}
\label{fig:excl}
\end{figure}
Finally, we have investigated the applicability of models trained on a given benchmark to the other benchmark datasets. 
The ability to train a model on a certain benchmark and apply it across the mass spectrum would be advantageous as the BSM model at hand features a largely unconstrained two-dimensional parameter space of possible masses for the smuon and neutralino. It would be preferable to be able to train a BDT model on one benchmark, and apply it across a large mass range, or indeed across all mass combinations. This would be subject to the similarities in features across the benchmarks. One may expect significantly different kinematic behaviour across mass separations, as the available phase space will modify final momenta and angular distributions. It might be expected then that a model trained within a certain mass regime would only produce similar results on benchmarks with a similar mass ratio $m_{\tilde{\mu}}/m_{\tilde{\chi}}$.
To test this, we train an {\sc XGBoost} model on a given benchmark ($m_{\tilde{\mu}}$, $r$), and apply that model to the other benchmarks. We have used the \textit{F-score} for training, and applied the $ams$ subsequently.\

The results are presented in figure~\ref{fig:excl}, where as before the training benchmark is encircled in red. Good performance is achieved when training on a benchmark within the yellow region and testing on that same benchmark. In that case, $2\sigma$ is achieved, demonstrating equivalence with the performance of the cut-and-count analysis. It is also clear that lower smuon masses perform better in general; even for training on higher masses, good results are achieved for lower masses. This is partially linked to the higher event numbers available in those regions due to larger cross sections, making those areas more accessible. It is notable that the lower smuon masses were not covered by the $2\sigma$ band of the cut-and-count, but appear achievable here. We also draw the reader's attention to the similarity in behaviours in benchmarks with a given smuon mass. As an example, the model trained on $(m_{\tilde{\mu}}, m_{\tilde{\chi}}) = (500,95)~{\rm GeV}$ displays good performance for both points featuring the same smuon mass. Moreover, the compressed region remains out of reach, even if training is done on a compressed benchmark. Similarly, the models which are trained on the \textit{compressed} cases show poor performance on other benchmarks as well as on themselves. This is visible in the top right subplot in figure~\ref{fig:excl}.\

From these results it is clear that there is some level of model generalisability. The \textit{compressed} regimes are entirely out of reach, although they may be reachable using searches exploiting the mono-jet and mono-photon topology~\cite{Birkedal:2004xn, Feng:2005gj}. In the case at hand,  \textit{neutral} and \textit{separated} regimes are accessible. However, the performance of the \textit{neutral} regime for the smuon masses degrades as the mass of the smuon in the training benchmark increases. The results suggest that the lower mass neutral benchmarks are accessible at experiments, and that they may all be investigated through a model trained on only one benchmark. This may be linked to the larger cross sections of lighter scenarios leading to more events in the detector.

\section{Summary and outlook}
\label{sec:summary}

In this work we have outlined the application of a BDT algorithm to a generic particle physics problem. We have outlined the key concepts in such a study, including a comparison of the widely used {\it auc} metric to the \textit{F-score} metric, where the latter may better handle the imbalanced datasets characteristic of particle physics problems. Indeed, we find that the \textit{F-score} is a more stable choice of metric and achieves better results. An in-depth feature exploration and analysis was performed, where a PCA was used in conjunction with a Shapley analysis and a feature permutation tool in order to assess the utility of the features at hand. In such an analysis this step is critical in understanding the data at hand in order to use its features correctly in an analysis.

We have compared the use of sophisticated features in previous cut-and-count analyses with respect to machine learning applications, and found that they may not be necessary in the given scenario. This outlines where machine learning may be used to simplify analyses, where the need for complicated feature construction may be lessened. This decreased dependence on the sophisticated ratio features is likely due to the way in which a BDT treats events which do not pass initial selections, constructing later trees to readdress the outlying points. In this way, simpler variables may be used in a number of ways with varying thresholds to carefully separate signal from background.\

We have also investigated the generalisability of a trained model, as BSM physics models such as the one studied here feature a large spectrum of possible masses to be scanned over. We have identified that such a scan would in principle be possible, but that the benchmarks should be sorted into regimes which dictate their kinematics. We have also identified that training on lower masses yields the best results and is the most compatible with a parameter scan. The machine learning model is however unable to generalise to higher mass regions, likely due to paucity of events there. Additionally, \textit{compressed} regions are out of reach and require a different analysis strategy.

In particular, we highlight that we are able to extend the exclusion region of Ref.~\cite{Fuks:2019iaj} to lower mass cases, which positively improves on the previous result. In addition, training on points in the sensitive region and testing on those same points yields results at least equivalent to those of the cut-and-count, indicating a consistency baseline. \

This work has acted as a 'proof of concept', and has highlighted many areas in which machine learning tools perform well and may, in some cases, out-perform cut-and-count analyses. Having paved the way, this work may now be extended to address more complicated physics collider signals, where difficult final states may benefit from such a tool. In addition, the extension of this analysis to involve higher order event generation would be a further test of the application of BDTs to a realistic collider example. While this work has focused only on BDTs, one may also consider additional machine learning techniques in the application the BSM physics analyses of this nature.


\section*{Acknowledgements}
The authors would like to thank Jack Araz for useful discussions. ASC is supported in part by the National Research Foundation of South Africa (NRF). LM is supported by the UJ GES 4IR initiative and thanks Nikita Kazeev for useful discussions. GEH is supported by UJ and  would like to thank Adrian Snyman for assistance with running the codes on their servers.

\bibliographystyle{utphys}
\bibliography{ppmlbib}

\end{document}